\documentclass[10pt,sigconf]{acmart}
\usepackage[utf8]{inputenc}
\usepackage{subcaption}
\usepackage[font=bf]{caption}
\usepackage{amsfonts}
\usepackage{booktabs}
\usepackage{amsmath}
\usepackage{color}
\usepackage{makecell}
\usepackage{bbding}
\usepackage{pifont}
\usepackage{wasysym}
\usepackage{amssymb}
\usepackage{algorithm}
\usepackage[noend]{algpseudocode}
\usepackage{balance}
\usepackage{enumitem}

\usepackage{url}

\usepackage{breakurl}

\newcommand{\squishlist}{\begin{itemize}[itemsep=1pt,parsep=2pt,topsep=3pt,partopsep=0pt,leftmargin=0em, itemindent=1em,labelwidth=1em,labelsep=0.5em]}
\newcommand{\squishend}{\end{itemize}}
\newcommand{\squishenum}{\begin{enumerate}[itemsep=1pt,parsep=2pt,topsep=3pt,partopsep=0pt,leftmargin=0em,listparindent=1.5em,labelwidth=1em,labelsep=0.5em]}
\newcommand{\squishsubenum}{\begin{enumerate}[itemsep=1pt,parsep=2pt,topsep=0pt,partopsep=0pt,leftmargin=0em,listparindent=1.5em,labelwidth=1em,labelsep=0.5em]}
\newcommand{\squishenumend}{\end{enumerate}}

\newcommand{\red}[1]{\textcolor{black}{#1}}

\newcommand{\xref}[1]{\S\ref{#1}}

\newcommand{\name}{Surface MIMO}

\renewcommand\footnotetextcopyrightpermission[1]{} 
\begin{document}


\title{Surface MIMO: Using Conductive Surfaces For MIMO Between Small Devices}

\author{Justin Chan, Anran Wang, Vikram Iyer and Shyamnath Gollakota}
\affiliation{University of Washington}
\email{{jucha, anranw, vsiyer, gshyam} @uw.edu}







\begin{abstract}
As connected devices continue to decrease in size, we explore the idea of leveraging everyday surfaces such as tabletops and walls to augment the wireless capabilities of devices. Specifically, we introduce {\name}, a technique that enables MIMO communication between small devices via surfaces coated with conductive paint or covered with conductive cloth. These surfaces act  as an additional spatial path that enables MIMO capabilities without increasing the physical size of the devices themselves. We provide an extensive characterization of these surfaces that reveal their effect on the propagation of EM waves. Our evaluation shows that we can enable additional spatial streams using the conductive surface and achieve average throughput gains of  2.6--3x for small devices. Finally, we also leverage the wideband characteristics of these conductive surfaces to demonstrate the first Gbps surface communication system that can directly transfer bits through the surface at up to 1.3~Gbps.
\end{abstract}




\fancyhead{}

\maketitle
\vskip -0.15in

\section{Introduction}
\label{sec:intro}

Multi-input and Multi-Output (MIMO) antenna technology was introduced into the Wi-Fi standard as part of 802.11n~\cite{80211n}. MIMO significantly increases Wi-Fi's achievable data rates using multiple antennas at the transmitting and receiving devices. However, these gains are difficult to achieve on many modern mobile devices like  phones and smart watches since their physical size intrinsically limits the number of antennas they can support. In fact, many of these mobile devices use only a single Wi-Fi antenna~\cite{single1,single2} and hence cannot achieve MIMO's multiplexing capabilities. 

We present {\it Surface MIMO}, a novel approach that enables MIMO between small devices, without the need to increase the physical size of the devices. Our key insight is that devices that are placed on the same surface (e.g., table) can exploit the surface itself as an additional spatial path for wireless signal propagation and thus achieve MIMO communication. Such a capability can enable two phones to rapidly share large files like photos or video by simply placing them on the same surface. As devices such as wearables, GoPros, and other smart objects continue to shrink in size while simultaneously capturing more data (e.g., 4k video), the ability to augment their wireless performance by simply placing them on a surface becomes an attractive option.

The fundamental problem however is that materials common to surfaces like walls and tabletops are not compatible with wireless communication. Specifically, wood, sheet rock and plastic are not conductive and cannot propagate radio signals.  Our approach instead is to augment everyday surfaces with a material that can facilitate wireless signal propagation. Specifically, we explore two minimally-invasive options: coating the surface with conductive paint and using conductive cloth over tabletops. Since re-painting walls is not uncommon and users regularly use tablecloths, curtains, and chair covers, such an approach could seamlessly integrate with existing furniture and surfaces, while requiring minimal aesthetic changes.

\begin{figure}[t]
    \includegraphics[width=.23\textwidth]{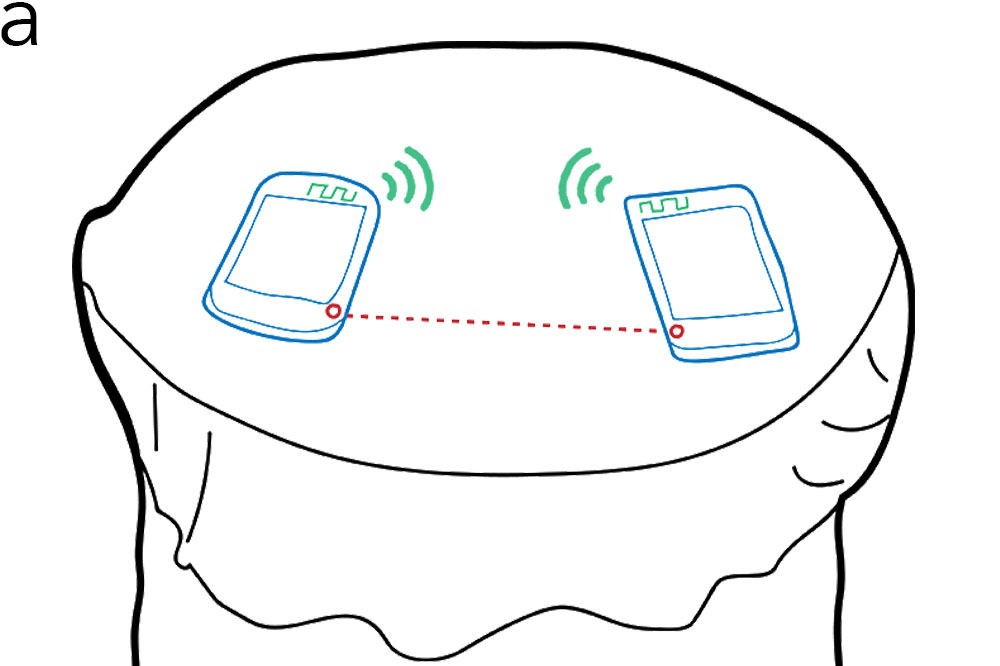}
    \includegraphics[width=.23\textwidth]{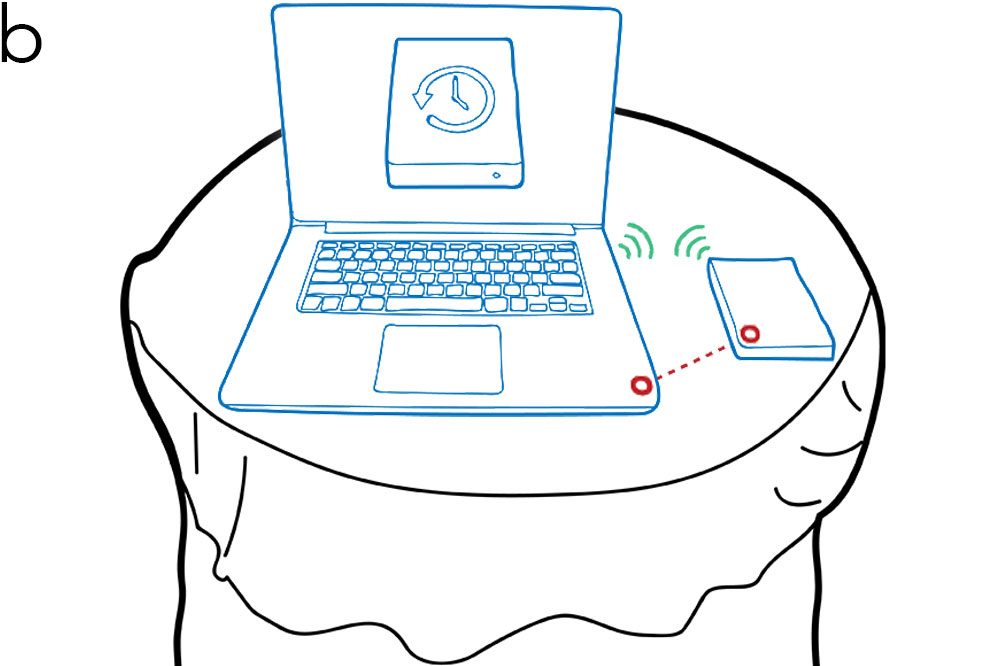}
\vskip -0.1in
    \caption{{\bf Surface MIMO communication.} \textmd{Surface MIMO uses conductive surfaces (e.g., tablecloth) as a second MIMO spatial stream and achieves MIMO communication capabilities. The separation between the antenna and the point of contact with the surface can be as small as 1~cm. \red{(a) Two single-antenna phones can leverage surface MIMO to rapidly share large files like photos or video by simply placing them on the same surface. (b) A laptop can rapidly back up or transfer files to a single-antenna external hard drive.}}}
    \label{fig:apps}
    \vskip -0.15in
\end{figure}

Surface MIMO leverages these conductive surfaces to enable MIMO capabilities. Specifically, similar to traditional RF communication,  surface MIMO uses a single antenna at the transmitting and receiving devices. In addition, as shown in Fig.~\ref{fig:apps}, it uses a single point of contact with the conductive surface to enable a second MIMO path {\it through} the surface. 

At a high level, the  propagation speed of EM waves through conductive materials is slower than their speed  in air. This propagation delay effectively creates a ``multi-path'' that has a different phase and amplitude, which allows us to enable MIMO capabilities on small devices. Specifically, traditional MIMO systems leverage multi-path to enable multiplexing and diversity gains, but require an antenna spacing that is a function of the wavelength. In contrast, in our approach, the propagation delay between in-air and over-the-surface transmissions, creates multiple propagation paths that are independent enough to enable MIMO gains even when the separation between the surface contact and the antenna is significantly smaller than half a wavelength.

\red{Note that conductive surfaces, while intuitively can be thought of as 2D wires, are not {shielded} and hence will radiate signals into the environment. This is unlike wired communication systems that use {shielding} to prevent cross-channel interference.} This results in non-diagonal surface MIMO matrices similar to traditional MIMO communication.

We present a detailed theoretical analysis and modeling of the surface MIMO channel and show that surface MIMO can be generalized beyond $2\times 2$ MIMO channel matrices. For example, by using two  points of contact on the conductive surface we can enable a $3\times 3$ MIMO communication channel. Intuitively, this is because of two key reasons: First, EM signals on the conductive surface experience significant multipath resulting from various reflections from impedance mismatches at the surface edges, unevenness of the material as well as various objects (e.g., books and laptops) placed on the conductive surfaces.  Second,  since the propagation of EM waves over conductive surfaces is slower than in-air, the contact separation required to leverage this multi-path for MIMO gains, is smaller than that required on the air. This allows the surface MIMO design to generalize to more than a single contact point on the conductive surface.

We empirically evaluate our surface MIMO design using Atheros AR9580 Wi-Fi cards and evaluate different $2\times 2$ and $3\times 3$ MIMO configurations. Our surface MIMO devices make contact with the surface using a small 1.6~mm diameter point. 
We measure the multipath properties of both conductive spraypaint as well as cloth. Our evaluation reveals that: 
\squishlist
\item $2\times 2$ and $3\times 3$ surface MIMO systems can achieve average end-to-end throughput gains of 2.6x and 3x over a single antenna Wi-Fi communication system. This shows that conductive surfaces can enable MIMO communication.
\item In comparison with a baseline MIMO system that uses an antenna separation of 6.25~cm, surface MIMO achieves 1.2x and 1.3x higher throughput for the $2\times 2$ and $3\times 3$ configurations, even when the point of contact is separated by only 1~cm from the Wi-Fi antenna. This shows that small devices can use surface MIMO to achieve MIMO communication.
\squishend

\begin{figure}[t]
    \includegraphics[width=.23\textwidth]{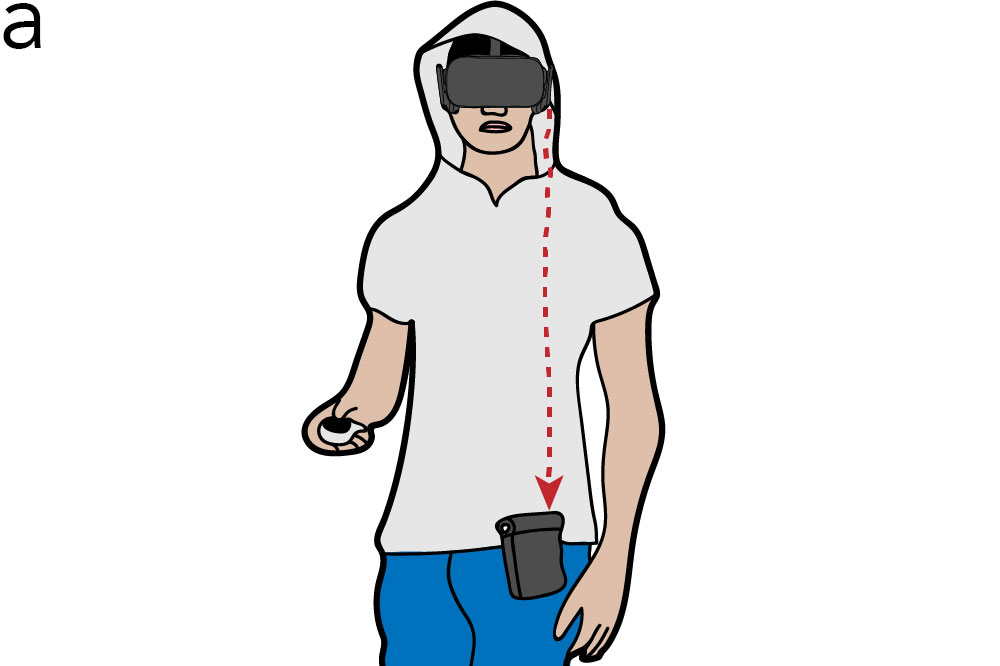}
    \includegraphics[width=.23\textwidth]{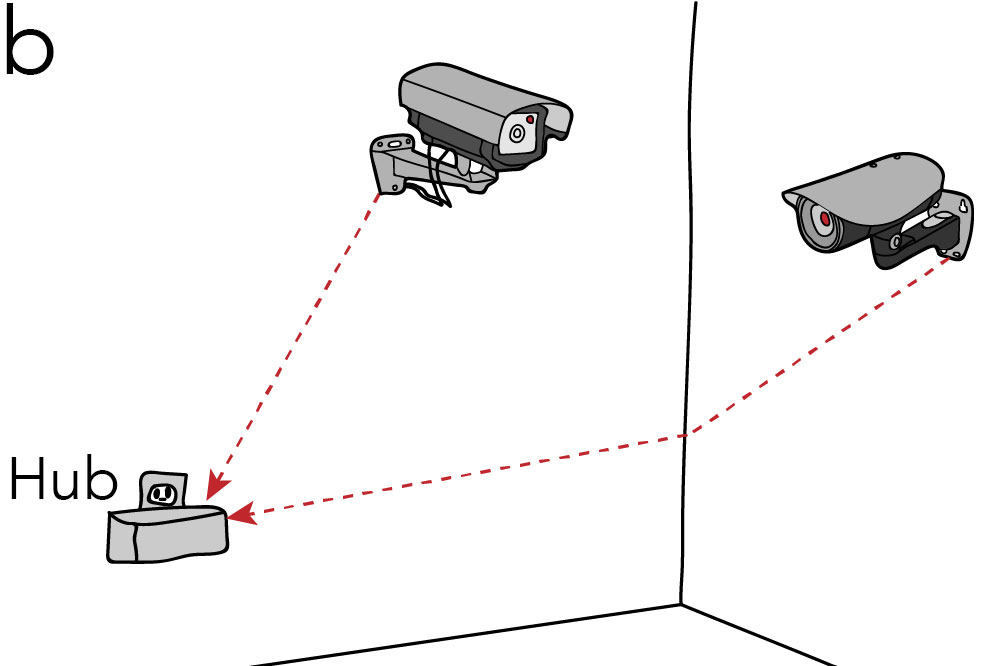}
\vskip -0.1in
    \caption{\red{{\bf Gbps surface communication. \textmd{The wideband nature of conductive surfaces can be leveraged to support Gbps communication. (a) A virtual reality headset streams video through a hoodie's conductive cloth to a wearable pack. (b) HD cameras on walls coated with conductive paint stream video to a plugged-in hub  through the wall.}}}}
    \label{fig:apps2}
    \vskip -0.1in
\end{figure}

{{\bf Gbps surface communication.} While exploring the properties of conductive surfaces for {\it Surface MIMO}, we find that we can also harness conductive surfaces to support  Gigabit per second data rates between devices placed on a surface, without using any traditional antennas. Specifically, we observe that these materials support wideband transmissions which is promising for achieving high data rates. The challenge, however, is that unlike  wired  systems (e.g. Ethernet) where the copper cable is \red{shielded} to  avoid  unintentional  radiation,  our  conductive  surfaces are not \red{shielded}, and hence radiate some RF signals  into  the  environment.  Thus,  if  we  were  to  implement an Ethernet-like system that uses sub--100~MHz frequencies over our uninsulated conductive surfaces, it would be incompatible with FCC regulations.} {Instead, we leverage the wideband nature of these conductive surfaces to achieve high data rate communication by simultaneously transmitting in the 900~MHz, 2.4~GHz and 5~GHz ISM bands.} \red{A key advantage of this design is that the signals radiated into the environment by these surface are 13--25~dB weaker than when using antennas. Thus, devices that share a surface can communicate with each other at Gbps bit rates, while minimizing the resulting interference in the radio environment.}

{This technique extends the utility of conductive surfaces beyond the form factor limited devices discussed above. For example, a device like a laptop could leverage these Gbps rates to rapidly back up or transfer files to an external hard drive on the table.} To demonstrate this capability we build a prototype of our wide-band hardware platform using off-the-shelf hardware components and show that we can support up to 1.3 Gbps physical-layer rates.

{\bf Contributions.} 
We make the following contributions.
\squishlist
\item We introduce the concept of surface MIMO that enables MIMO communication between small devices via conductive surfaces. Further, we identify and characterize conductive spraypaint and fabric cloth as mechanisms to augment everyday surfaces with conductive properties.
\item We provide a theoretical model as well as a detailed end-to-end experimental characterization of our surface MIMO channel using commodity Wi-Fi hardware.
\item Finally, we present the first communication design that can support Gbps data rates over surfaces. We prototype our design with off-the-shelf hardware and demonstrate the feasibility of Gbps speeds over conductive surfaces.
\squishend

\red{{\bf Applications.} Our surfaces can be thought of as a medium for communication between devices that are close to each other similar to NFC. However unlike NFC which has a low data rate of 424~kbps, we achieve orders of magnitude higher bit rates and hence can enable a wider range of wireless applications. In addition, these high bit rates can enable wireless cameras that transmit over conductive walls as well as VR headsets where conductive cloth can be used for Gbps communication to a wearable pack~\cite{magicleap} as shown in Fig.~\ref{fig:apps2}.}
\section{Related Work}
\label{sec:related}
There has been significant work in the wireless community on MIMO communication~\cite{mimo1,mimo2,mimo3,mimo10,mimo11,mimo12,mimo13}. We focus this section, on the related work on conductive surfaces. 

{\bf Surface Networking.}  \cite{pushpin-misc0} characterized the use of a single conductive surface made of paper for communication by analyzing the surface resistivity and received signal strength for surfaces of different dimensions. They achieve a data rate of 100~kbps at a distance of 2~m. While our approach also uses a single conductive surface, our surfaces are made from conductive paint and cloth which can be applied to any common surface that may not have perfectly rectangular dimensions. More importantly, we introduce the concept of surface MIMO that has not been explored in prior work. Furthermore, to the best of our knowledge, we show for the first time, Gbps data rates on conductive surfaces.

Pushpin based designs use specially constructed multi-layer surfaces~\cite{pushpin0,pushpin1,pushpin-misc1,pushpin-misc2}. Specifically, it uses two conductive sheets that are separated by an insulation layer (e.g., rubber). The pushpin on the device creates a hole through the insulation layer to make contact with the conductive sheets so as to transfer power and data. ~\cite{japan1,japan2,japan3} uses an insulation layer to separate a metallic sheet and a specially designed metal sheet with a mesh pattern. This particular pattern allows the devices to be placed in contact with the surface using a contact that is the size of a Wi-Fi antenna, but without pushpins. In contrast to these approaches, we focus on data transmission on a single conductive layer which have several advantages: 1) it is simpler to fabricate and only requires a single point of contact between the device and the surface 2) short-circuits will not occur due to accidental contact between signal and ground planes. 

The networked surface project~\cite{surfacenetwork} strategically places conductive tiles in specific patterns and has objects connect to the surface using circular pads that are designed to map specific connection points onto these tiles. This approach is harder to fabricate as it requires large pads and carefully managing the large number of tiles as well as placement on specific connection points. Despite using special surfaces, pushpin~\cite{pushpin0} and networked surfaces~\cite{surfacenetwork} use lower frequency signals that do not achieve Gbps speeds and further radiate into the environment which can be problematic for FCC compliance when targeting higher-data rates. Our paper instead introduces the concept of surface MIMO which shows multiplexing and diversity gains for small devices using conductive surfaces. Furthermore, we show how to achieve Gbps bit rates on conductive surfaces using the ISM frequency bands, which being compatible with FCC.

{\bf Printing antennas.} \cite{highly} print antennas resonant at RF frequencies for flexible substrates like paper and fabric.~\cite{darpa} spray paints antennas using stencils on flexible materials like fabrics.~\cite{goog} uses a custom-designed spray paint containing numerous capacitors to create antennas. These designs are focused on creating antennas for over-the-air communication, i.e., as a printed substitute for antennas used in traditional wireless communication over the air. However, our focus is on communication on the surface itself.

{\bf Rapid prototyping and eTextiles.} Conductive fabrics and inks have been used as wires for smart fabric designs like the Lilypad Arduino~\cite{lilypad} and electronic designs like the CircuitScribe~\cite{circuitscribe} on paper. Conductive spray-paints have been used for prototyping circuit boards~\cite{sketching,micro,circuit,paperpulse,pulp,chibitronics}, capacitive sensors~\cite{electrick,instant,printsense,sensortape,tactiletape,leveraging,modular,midas,handcrafting} and printing RFID antennas using stencils~\cite{paperid}. In contrast our paper focuses  on the MIMO communication capabilities provided by conductive surfaces.

\section{Surface MIMO}
\red{It is well known that for conventional MIMO communication over the air, there should be at least a separation of half a wavelength (around 6cm for 2.4GHz ISM band) between antennas on each device. How is MIMO between small devices theoretically possible with the help from a shared conductive surface? } In this section, we provide a theoretical analysis of conductive surfaces as a communication medium by modeling their propagation characteristics. We then extend this analysis and show how it can be used for MIMO.

\subsection{Formal Channel Definitions}

We define a single-input single-output communication channel utilizing a conductive surface as follows: Say, a transmitter $TX$ is communicating with a receiver $RX$. At least one of them has a conductive contact touching the shared conductive surface, while the other, either has another conductive contact on the surface or a normal antenna in contact with the air. The conductive surface has conductivity $\sigma$, permeability $\mu$, and size $W\times H$, with negligible thickness. We consider a rectangular surface for simplicity but the analysis can be extended to other surface shapes. Finally, the transmitter and receiver do not share an explicit common ground; we therefore model them as being weakly capacitively coupled to a common low potential, or ``earth ground''. 

\begin{figure*}
\includegraphics[width=\textwidth]{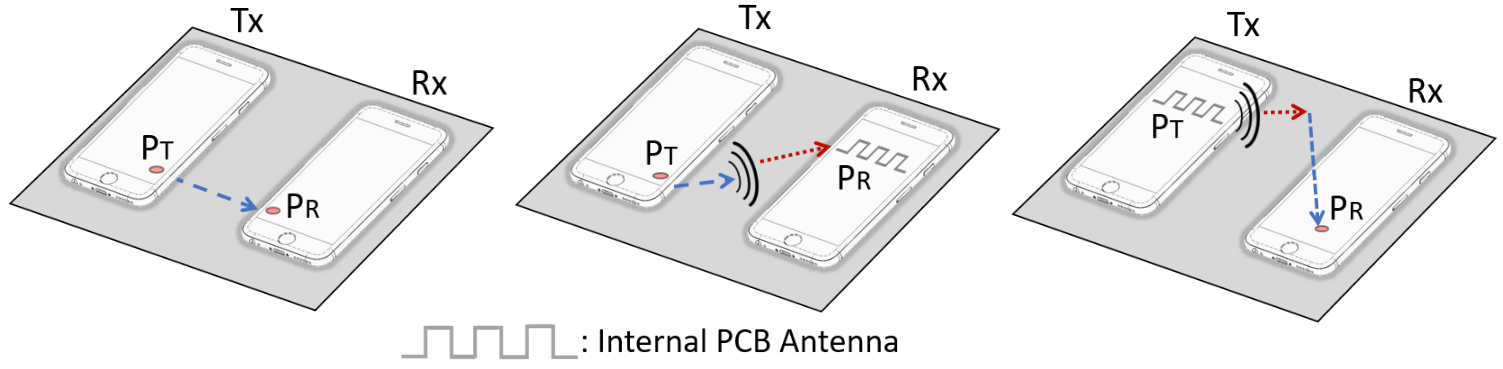}
\caption{Surface communication topologies. \textmd{(a) Surface-surface (b) surface-air (c) air-surface. Blue lines represent EM waves being transmitted across the surface. Red lines represent EM waves being transmitted in the air.}}
\label{fig:ill_channels}
\end{figure*}

We define three different channels based on whether $TX$ and $RX$ are touching the surface (Fig.~\ref{fig:ill_channels}).

\begin{definition}
\textbf{Surface-surface channel}. 
The communication channel from $TX$ to $RX$ where both $TX$ and $RX$ are in touch with the surface $\mathbb{S}$. Specifically, relative to the top left corner of the surface, the $TX$ is placed at position $P_{TX}=(X_{TX}, Y_{TX})$ and the $RX$ is placed at position $P_{RX}=(X_{RX}, Y_{RX})$.  
\end{definition}
\begin{definition}
\textbf{Surface-air channel}. 
The communication channel from $TX$ to $RX$ where $TX$ is touching the surface $\mathbb{S}$, placed at position $P_{TX}=(X_{TX}, Y_{TX})$ relative to the top left corner of the surface, while $Rx$ has a normal RF antenna, located at position $P_{RX}=(X_{RX}, Y_{RX}, Z_{RX})$. 
\end{definition}
\begin{definition}
\textbf{Air-surface channel}. 
The communication channel from $TX$ to $RX$ where $RX$ is touching the surface $\mathbb{S}$, placed at position $P_{RX}=(X_{RX}, Y_{RX})$, while $TX$ has a normal RF antenna, located at position $P_{TX}=(X_{TX}, Y_{TX}, Z_{TX})$. 
\end{definition}

\subsection{Signal Propagation on Surfaces}

{In our surface communication system, a transmitted signal can propagate through the surface itself or radiate from the surface into the air. Additionally, signals traveling over the air can also be absorbed by the surface.} In this subsection, we provide an overview of each of the propagation paths.

\subsubsection{Surface-surface channel.} When the transmitter and receiver are in contact with the surface they effectively form a circuit. While the transmitter and receiver do not share a common ground, there still exists a weak electric field between the exposed ground terminals and earth, which can be modeled as a small capacitor that provides a return path~\cite{onbody}.
This communication path through the conductive surface can be modeled as a complex circuit depending on its material, size and shape. We approximate the path through the surface as a transmission line traditionally modeled by an RLC circuit. As a whole, the surface introduces attenuation that varies as a function of distance as some of the incident power is dissipated into it. 

Prior work has physically characterized EM propagation through conductive transmission lines~\cite{matick2000transmission}. In particular, within RF  bands, the signal attenuates exponentially over distance. In comparison, EM waves that travel through non-conductive materials like air do not in theory lose energy over distance, but rather continue to spread in space resulting in lower received power density versus distance. 

More formally, using Maxwell's equations~\cite{griffiths2017introduction}, we can characterize the surface as introducing some attenuation in amplitude and change in phase versus distance through the transmission line. For example, when the transmitter sends $V(t)=V_{in}\exp(j\omega t)$, the received signal received at distance $d$ can be represented as $V_{in} \exp(-\alpha d)\exp(j\omega t-j\beta d)=V(t)\exp(-\alpha d-j\beta d)$ where $\alpha$ and $\beta$ are constants determined by the material properties of the surface such as conductivity. 

Because properties of surfaces such as conductivity and skin depth are functions of frequencies, high frequency signals experience greater attenuation\cite{gibson2010channel}. As a result, extremely high frequencies like millimeter waves do not propagate well over conductive surfaces and dissipate as heat on the surface.\footnote{Our conductive surfaces are in fact not compatible with 802.11ad frequencies. This is because, at such high frequencies, the signal not only attenuates significantly over distance, but also radiates a non-negligible amount of heat on these surfaces. In fact, our experiments with a 60~GHz MIMO system on the surface fried the CPUs of our testbed equipment and began to heat up our surface.} Next, consider a 2D conductive surface. Assuming the signal propagates omni-directionally, the received signal at distance $d$ from the transmitter can be represented as, 
\begin{equation}
V_{Rx}[t]=V_{Tx}[t]\exp(-\alpha d-j\beta d)\frac{d_0}{d}
\label{eq:surface_att}
\end{equation}

Further, the effect of the transmission line on the phase of the signal is captured by the $\beta$ term. The propagation speed (phase velocity) of the signal inside conductive materials, which can also be derived from the above equation, is
$v\propto \frac{\omega}{\beta}$. Note that this is slower than the speed of light in air under RF frequencies.

\subsubsection{EM wave propagation over the air} The energy that EM waves within the surface lose over distance is converted into heat as well as radiates emissions into the air.  The EM waves that propagate over the air travels at the speed of light and follows the inverse-square law attenuation model\cite{gibson2010channel}, like other wireless communications such as Wi-Fi. Formally, if we assume the EM radiation is omni-directional, the received signal at distance $d$ can be represented as
\begin{equation}
V_{Rx}[t]=V_{Tx}[t]\exp(-j\omega\frac{d}{c})\frac{d_0^2}{d^2}
\label{eq:air_att}
\end{equation} 
where $d_0$ is a constant and $c$ is the speed of light.

\red{To conclude, the signal propagation and attenuation through the surface are quite different from those through the air. Next, we show that the difference can be utilized to achieve MIMO communication for small devices.}

\subsection{Modeling the Surface Channel}\label{sec:model_channel}
By combining the above propagation models together, we can derive the surface channel model for a particular frequency band. 
Based on the transmission line model we describe, the dominant path for low frequency signals is through the surface itself. However at higher frequencies (\textit{i.e.}, 2.4GHz), the radiation to the air dominates. 

Specifically, the EM waves from the transmitter first propagate through the surface. At a point $p_1$, a portion of the remaining signal is emitted into the air. At the receiver side, at a point $p_2$, radiated EM waves in the air may be absorbed into the surface the surface, combined with the residual EM waves through the surface, received by the receiver. 

We assume the above transformations only happen once between the transmitter and receiver to simplify our model.

Formally, we let $x[t]$ and $y[t]$ denote the complex transmitted signal from $TX$ and received signal from $RX$ at time $t$ respectively. Assuming no multipath, we derive the following equations by splitting the propagation into segments from $P_{TX}$ to $p_1$, $p_1$ to $p_2$ and $p_2$ to $P_{RX}$.

Specifically we derive a model for the surface-surface channels based on these intermediate positions as: 

\begin{align}\begin{split}
y[t]=& C_1\iint_{p_1(x_1, y_1)\in\mathbb{S}}\iint_{p_2(x_2, y_2)\in\mathbb{S}}A_{\mathbb{S}}(d_1)A_{\mathbb{S}}(d_3) A_{air}(d_2)x[t] \\
& dx_1dy_1dx_2dy_2 +{n(t)} \\
=& H_{SS}x[t]+n(t)
\end{split} \label{eq:ss_channel}
\end{align}

where $d_1=|P_{TX}-p_1|$, $d_2=|p_1-p_2|$, $d_3=|P_{RX}-p_2|$, $A_{\mathbb{S}}(\cdot)$ is the attenuation factor model on the surface (Equation~\ref{eq:surface_att}), $A_{air}(\cdot)$ is the attenuation factor model in the air (Equation~\ref{eq:air_att}), ${n(t)}$ is Gaussian noise and $C$ is a constant.   The above integral captures the fact that the EM waves from the transmitter can first propagate through the surface to a distance $d_1$, radiate into the air for a distance $d_2$ and propagate in air and then propagate on the surface again for a distance of $d_3$. 

The received signal for surface-air channels can similarly be written as,

\begin{align*}
y[t]=& C_2\iint_{p(x, y)\in\mathbb{S}} A_{\mathbb{S}}(|P_{TX}-p|) A_{air}(|p-P_{RX}|)x[t] dxdy + n(t) \\
=& H_{SA}x[t]+n(t)
\end{align*}
The equation for the air-surface channel is symmetric to the above channel equation:

\begin{align*}
y[t]=& C_3\iint_{p(x, y)\in\mathbb{S}} A_{air}(|P_{TX}-p|) A_{\mathbb{S}}(|p-P_{RX}|)x[t] dxdy + n(t) \\
=& H_{AS}x[t]+n(t)
\end{align*}

\red{Note that $H_{SA}$ and $H_{AS}$ are largely uncorrelated because of the difference of $A_{air}$ and $A_{\mathbb{S}}$.  }

\subsection{Surface MIMO Channel}

At high SNRs, MIMO {multiplexing}, can be used to improve the channel capacity by $N$ times. Specifically, each signal received by the $i$th receive antenna is a linear combination of signals from each of the transmit antennas:
\begin{equation*}
\mathbf{y}[t]=
\begin{bmatrix}
y_1\\
y_2\\
\cdots\\
y_N
\end{bmatrix}
=
\begin{bmatrix}
H_{11} & H_{12} & \cdots & H_{1N} \\
H_{21} & H_{22} & \cdots & H_{2N} \\
\cdots & \cdots & \cdots & \cdots \\
H_{N1} & H_{N2} & \cdots & H_{NN} \\
\end{bmatrix}
\begin{bmatrix}
x_1\\
x_2\\
\cdots\\
x_N
\end{bmatrix}+\mathbf{n}(t)
\end{equation*}
where $\mathbf{H}$ is the \textit{channel matrix}.  It is well-known that  when all $N^2$ spatial paths are independent from each other, we can achieve an $N$ fold capacity increase over $N$ spatial channels.

At low SNRs, we can use MIMO diversity to improve the {received signal's SNR}. For example, for a $N\times N$ MIMO channel, we can transmit the same signal with different phase offsets on all the transmit antennas, such that the receiver can receive an in-phase combinations of all the transmitted signals.  At the receiver side, we can align the received signals such that they are in-phase, and use maximal ratio combining (MRC) or diversity coding techniques to combine them into a single signal. When the $N^2$ different spatial paths between each transmit and receive antenna are independent, we can achieve as much as an $N^2$ increase in SNR. 

In this section, we analyze the MIMO channels that can be enabled by the use of conductive surface. 

\subsubsection{Single point of contact} We first consider the scenario where the transmitter $Tx$ and receiver $Rx$ each has regular antennas $Tx_A$ and $Rx_A$ respectively that communicate via air. In addition, they also use a single point of contact on a shared conductive surface, $Tx_S$ and $Rx_S$ touching the surface. Recall that when the RF transmission is solely over the air, ignoring multipath, the received signal over the air, $y_{AA}$, can be represented as,
${y_{AA}}[t]={H_{AA} x}[t]$. Here the channel over the air, $H_{AA}$ can be written as,
${H_{AA}}=A_{air}(|P_T-P_R|)exp(-j\omega \frac{|P_T-P_R|}{c})$, where $A_{air}(d)$ is the attenuation over air as a function of distance $d$ and $P_T$ and $P_R$ are the position coordinates for the transmitter and receiver.
Now combining this with the surface-surface ($H_{SS}$), surface-air ($H_{SA}$), and air-surface ($H_{AS}$) channels derived in~\xref{sec:model_channel},  the surface MIMO channel can be written as,
$$\mathbf{H}=\begin{bmatrix}
H_{SS} & H_{SA}\\
H_{AS} & H_{AA}\\
\end{bmatrix}$$
The above equation reveals the following. First, \red{Since $H_{AS}$, $H_{SA}$, $H_{AA}$, $H_{SS}$ are non-correlated, the above matrix has two non-zero eigenvalues and is well-conditioned.} 
Second, while the propagation on the surface  deteriorates with distance faster than that on air, at close by distances (within a few meters), the signal strength over the surface is strong enough to create a full-rank matrix (as is demonstrated in our evaluation). Note that since our application is for devices that are close by placed on a table, this distance is sufficient for practical application.
Third, the above equation shows that even in the absence of multi-path over the air, the signal propagation over the surface, provides an additional multi-path that can achieve a MIMO multiplexing gain.
Finally, since the independence of $H_{SS}$ and $H_{AA}$ is not dependent on the separation between the transmit (receive) antenna and the point of contact at the transmitter (receiver), the above design can potentially enable a MIMO system even when the separation is much smaller than half a wavelength.

\subsubsection{More than one point of surface contact} Next we describe scenarios where  the transmitter and receiver both have two contacts with the conductive surface. Firstly, based on Eq.~\ref{eq:ss_channel} we know that the surface can be thought of as a multipath channel due to the combination of the EM waves inside the surface and over the air. Moreover, EM waves traveling in the conductive surface bounce back at its boundaries due to impedance mismatches between the conductive surface and the air. These two reasons create a diverse multipath profile (see~\xref{sec:multipatheval}) on the surface, which is necessary for MIMO communication with multiple contact points.

More importantly, due to the slow propagation speed of EM wave in the conductive surface, the wavelength of the signal that is transmitted through the surface becomes shorter. Recall that the ideal antenna separation is half the wavelength. With this decreased wavelength, each contact only needs to be separated by a smaller distance to enable MIMO. Because of this we can enable $3\times 3$ or larger antenna-free MIMO systems on small devices by leveraging the multipath properties of the conductive surface.

\section{Experimental Evaluation}
\label{sec:char}

\red{While the theory is promising, in practice, the actual MIMO performance depends on various parameters including the surface material type and properties, the locations of devices, as well as the environment. To this end, }
we conduct extensive experiments to analyze the properties of the surface MIMO channel. We first analyze the  propagation over conductive surfaces. We then measure the multi-path profile of these surfaces in practice. Finally, we demonstrate the performance of surface MIMO using off-the-shelf Wi-Fi hardware.

\begin{figure}[t]
\centering
\begin{subfigure}[b]{0.44\textwidth}
\includegraphics[width=\textwidth]{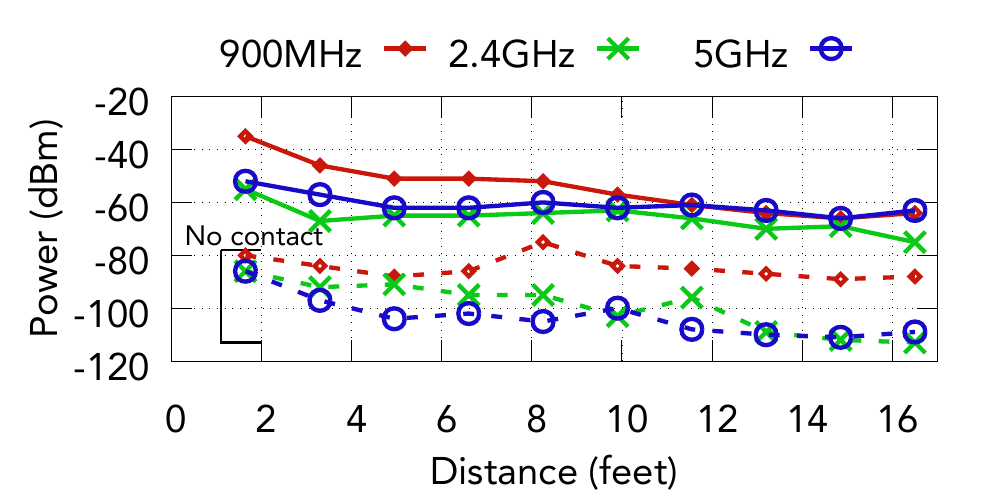}
\caption{\textmd{Conductive spraypaint}}
\end{subfigure}
\begin{subfigure}[b]{0.44\textwidth}
\includegraphics[width=\textwidth]{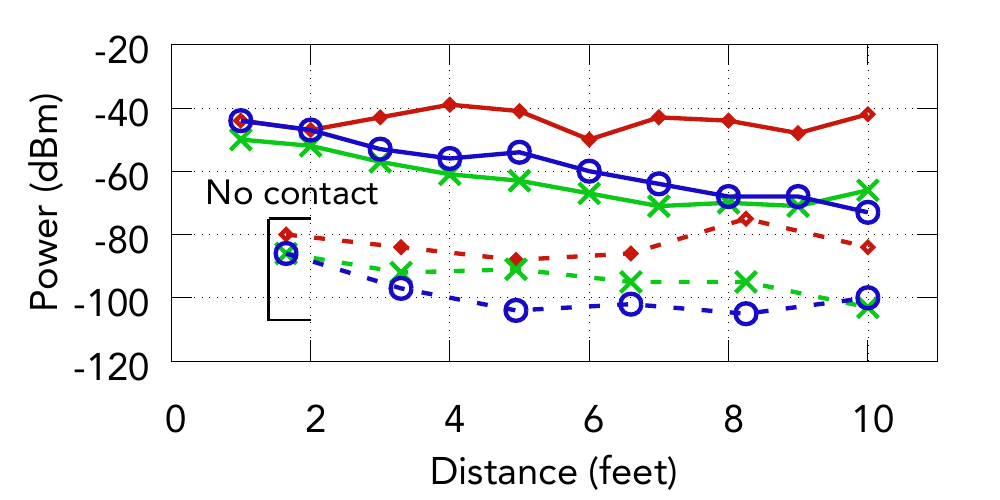}
\caption{\textmd{Conductive cloth}}
\end{subfigure}
\caption[caption]{{\bf Signal attenuation over distance \red{for conductive spraypaint and cloth.}} \textmd{In the absence of contact \red{ with the surface, the received power} is close to \red{the noise floor}, demonstrating that communication between the devices occur due to their contact with the conductive surface.}}
\vspace{-0.1in}
\label{fig:loss}
\end{figure}

\subsection{Surface Channel Characterization}
\label{sec:surfacechar}
We first understand how the signals attenuate over distance on the conductive surfaces. To run experiments over large distance on a surface, we paint a 16 by 2 feet long paper with conductive spraypaint~\cite{mgc}. Our conductive cloth testbed is constructed by sewing together ten pieces of conductive fabric~\cite{fabric} that measures 1 by 1 feet each. This results in a 10 by 1 feet tablecloth. We measure the attenuation across a range of frequencies from 900~MHz, 2.4~GHz to 5~GHz. These frequencies are sent using a USRP that makes contact using the center pin of an SMA connector which has a diameter of 1.6~mm. The transmit power is set to -3~dBm and our receiver uses a 5~GSps oscilloscope~\cite{scope5}. We attached a SMA connector to the oscilloscope and placed the connector in contact with the conductive surface. The USRP was plugged into a portable battery that does not have the same ground as the oscilloscope, which was plugged into the wall.

Fig.~\ref{fig:loss} plots the results for the two conductive surfaces. The plots reveal the following: The signal attenuation is similar for 900MHz, 2.4GHz and 5GHz, and higher frequencies attenuate a little faster than lower frequencies. This also shows that our conductive surfaces work across a very wideband of frequencies, and can be used for many different applications. For comparison, -70~dBm is a `good' Wi-Fi signal~\cite{goodwifi,goodwifi2}. Further, for conductive cloth, we can see that all three frequencies bands perform similarly to conductive spraypaint.  Finally, the signal strength when the transmitter and receiver are not in contact with the surface is close to noise. This demonstrates that the communication between the devices occurs due to their contact with the surface.

\begin{figure}[t]
  \centering
    \includegraphics[width=0.45\textwidth]{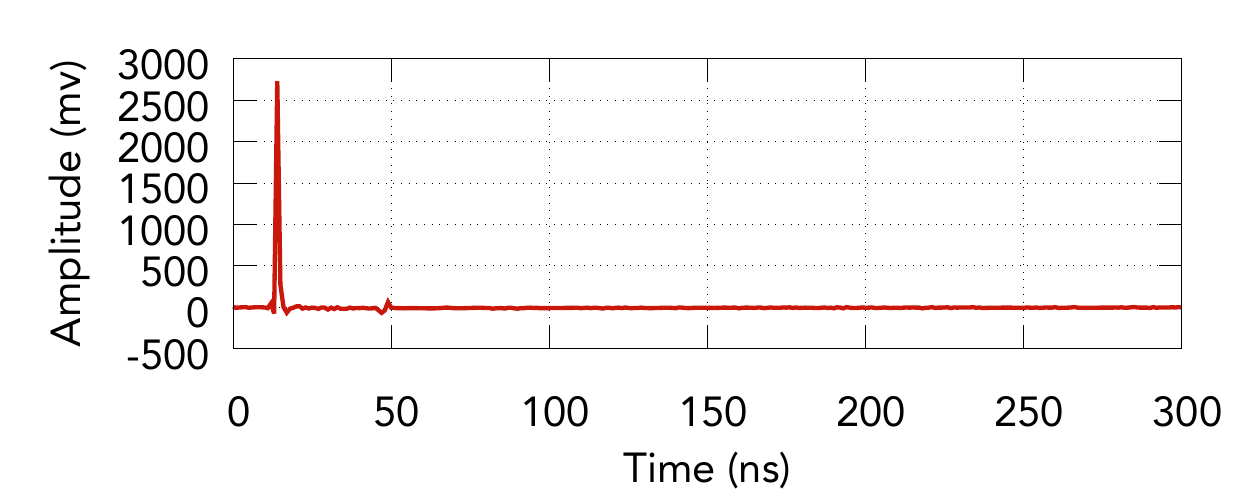}
\vskip -0.1in
  \caption[caption]{{\bf Baseline pulse.} \textmd{Sending a 1 nanosecond pulse, when transmitter and receiver are connected by a copper cable.}}
\vskip -0.2in
\label{fig:pulse}
\end{figure}

\begin{figure}[t!]
  \centering
    \includegraphics[width=0.23\textwidth]{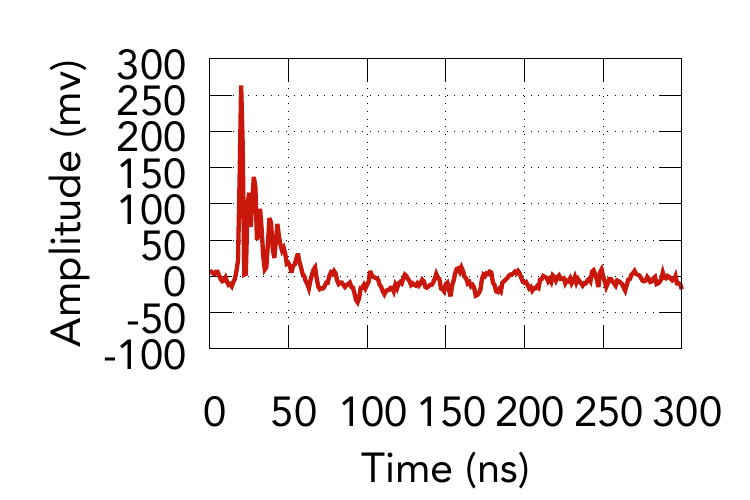}
    \includegraphics[width=0.23\textwidth]{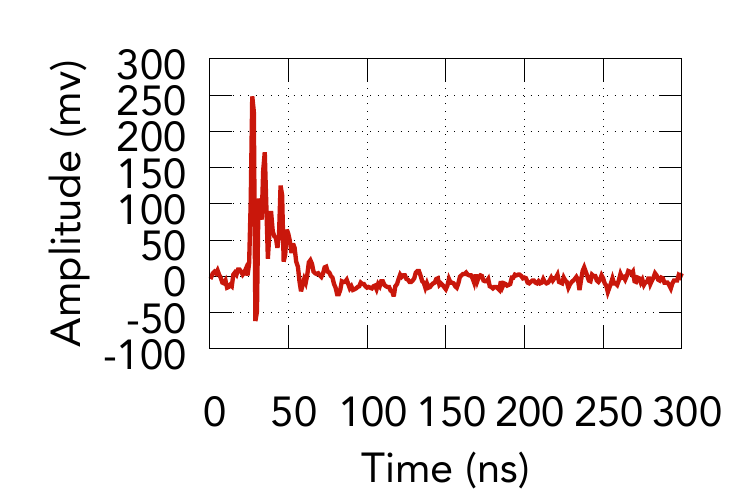}
    \includegraphics[width=0.23\textwidth]{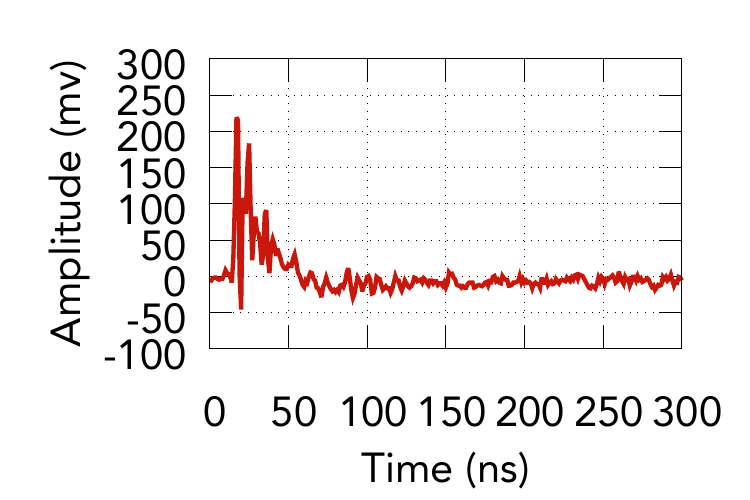}
    \includegraphics[width=0.23\textwidth]{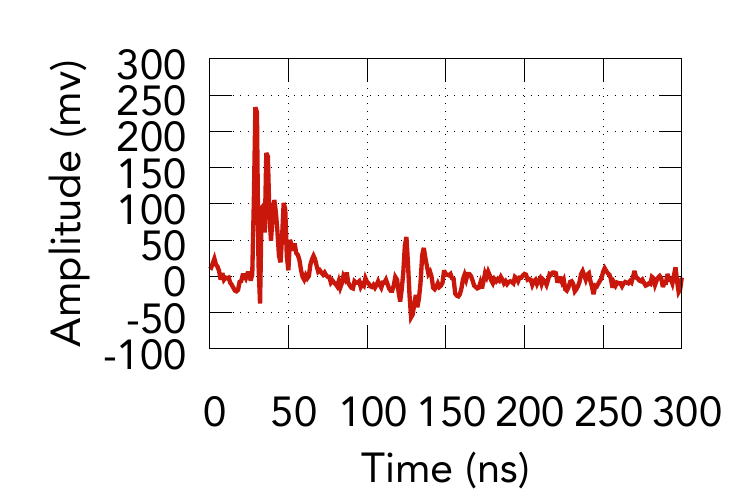}
\vspace{-0.1in}
    \caption[caption]{{\bf Multipath on cloth.} \textmd{When there is (a) no object (b) metal (c) plastic (d) wooden object on the tablecloth.}}
\label{fig:pulse-cloth}
\vspace{-0.1in}
\end{figure}

\subsection{Surface Multipath Propagation}\label{sec:multipatheval} We transmit a short one-nanosecond pulse through the conductive surface and record the received signal to determine its multipath properties. We use Analog Device's FMCDAQ2 \cite{daq2} which provides 1~Gsps DAC and ADC processing. We interface with the DAC/ADC using Xilinx's Kintex KC705 FPGA~\cite{kc705}. The DAC and ADC each make contact with the conductive surface using the 1.6~mm diameter center pin of an SMA connector. As a wired reference, we connect the DAC and ADC directly to each other using a coaxial cable and measure the received signal. 

Fig.~\ref{fig:pulse} shows the received pulse which is close to ideal and has no major reflections across the wire. Next, we conduct the same experiment across our tablecloth. We record a signal on the surface when there are no objects placed on the surface. Also to test whether objects on top of the surface affect multipath profiles, we place a metal box, a plastic box and a wooden plank respectively onto the surface and capture the received signal in each of these cases. Fig.~\ref{fig:pulse-cloth} show the captured signals. The figures reveal that the channel has significant multipath. \red{Additionally, the profile shows the amount of noise on the channel 300 nanoseconds after the initial pulse is sent.} While the placement of different objects on the surface slightly changes the multipath profile, it does not prevent a strong signal from being received.

\begin{figure}[t]
  \centering
    \includegraphics[width=0.37\textwidth]{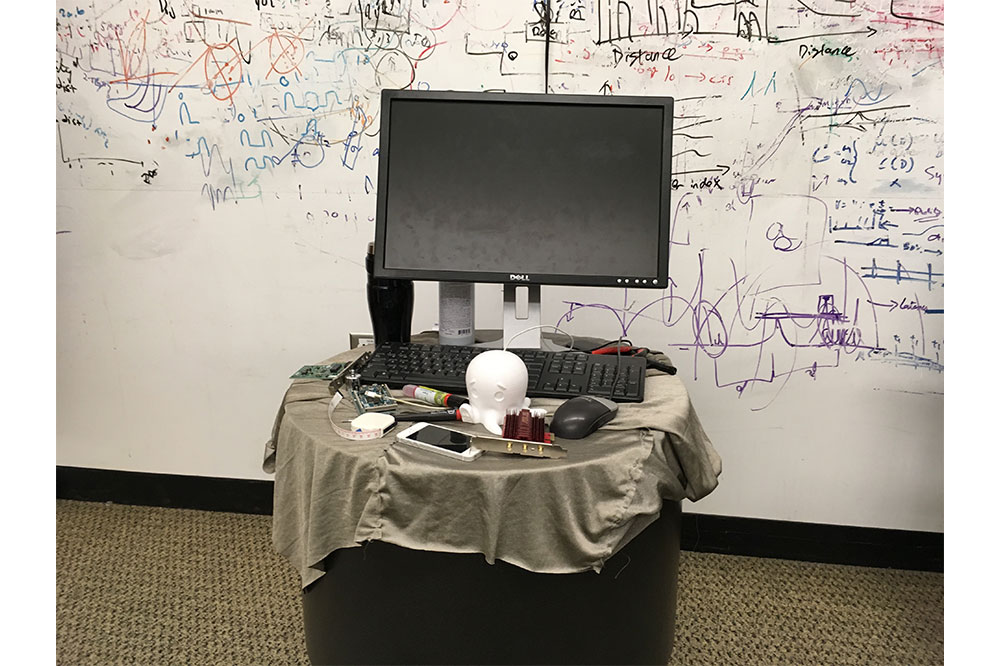}
    \includegraphics[width=0.23\textwidth]{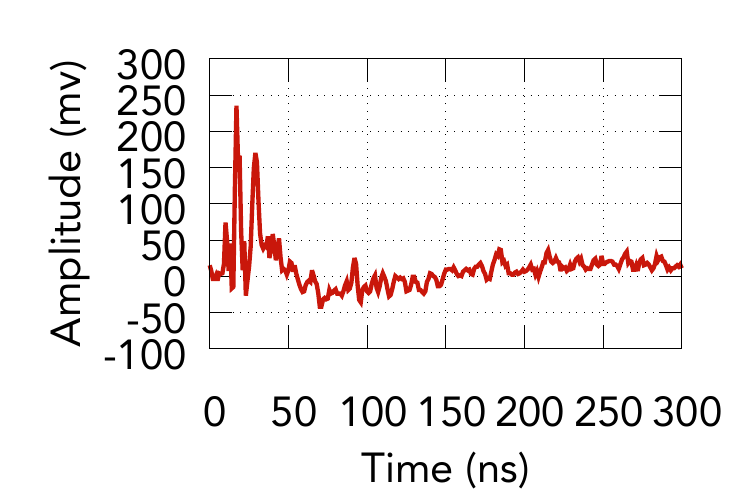}
    \includegraphics[width=0.23\textwidth]{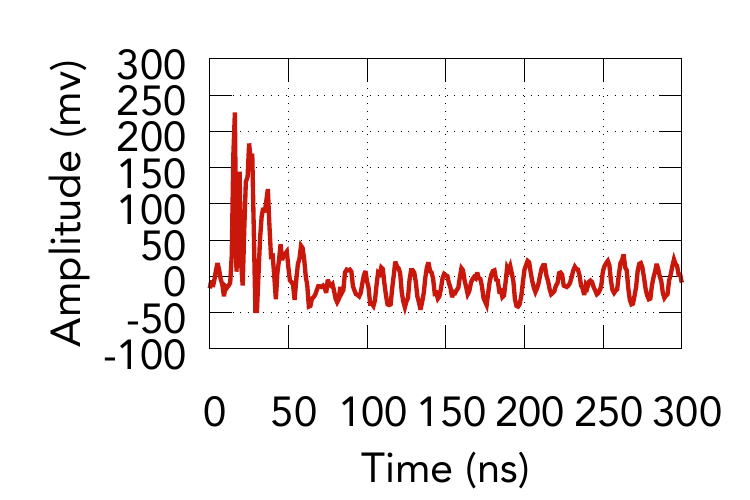}
\vspace{-0.1in}
  \caption[caption]{{\bf Busy tabletop.} \textmd{With (a) spraypainted sheet and (b) tablecloth. The receiver can still get a strong signal.}}
\label{fig:pulse-table}
\vspace{-0.2in}
\end{figure}

Finally we use the busy tabletop surface shown in Fig.~\ref{fig:pulse-table} with both the tablecloth and a 3 by 2 feet spraypainted sheet, which we believe is an extreme deployment scenario for these surfaces. Again, we send a short pulse through the surface and capture the signal. We observe similar multipath profiles and more importantly the receiver can still decode a strong signal. \red{We note that the multipath profile is similar even for different configurations of objects on the table top and when humans are in contact with the surface.}

\subsection{Evaluating Surface MIMO}
\label{sec:mimo}

\subsubsection{Experimental setup} 
We run experiments with Atheros AR9580 Wi-Fi cards~\cite{atheros} which support a $3 \times 3$ 802.11n configuration. We use these cards in a two-antenna mode where one of the ports at the transmitter and receiver are connected to Wi-Fi antennas. The second port is connected via a contact to the conductive surface. And the third port is disabled for by $2 \times 2$ MIMO experiments. We use a PCB trace antenna at 2.4~GHz which is meandered with a total length of 1.5~cm. We use these meandering antennas as they are common on small mobile devices like smartphones and other IoT devices~\cite{pcb2}. \red{In our short range experiments we find there is little difference in performance between PCB antennas and typical dipole antennas found on access points.}

\subsubsection{Surface MIMO channel}
We use the Atheros CSI tool~\cite{precise} to extract the channel state information (CSI) between the various antennas at the transmitter and receiver. We set the transmitter to use a 40~MHz bandwidth with a 400~ns OFDM guard interval at 2.4~GHz. 

\begin{figure}[t]
  \centering
    \begin{subfigure}[b]{0.47\textwidth}
    \includegraphics[width=\textwidth]{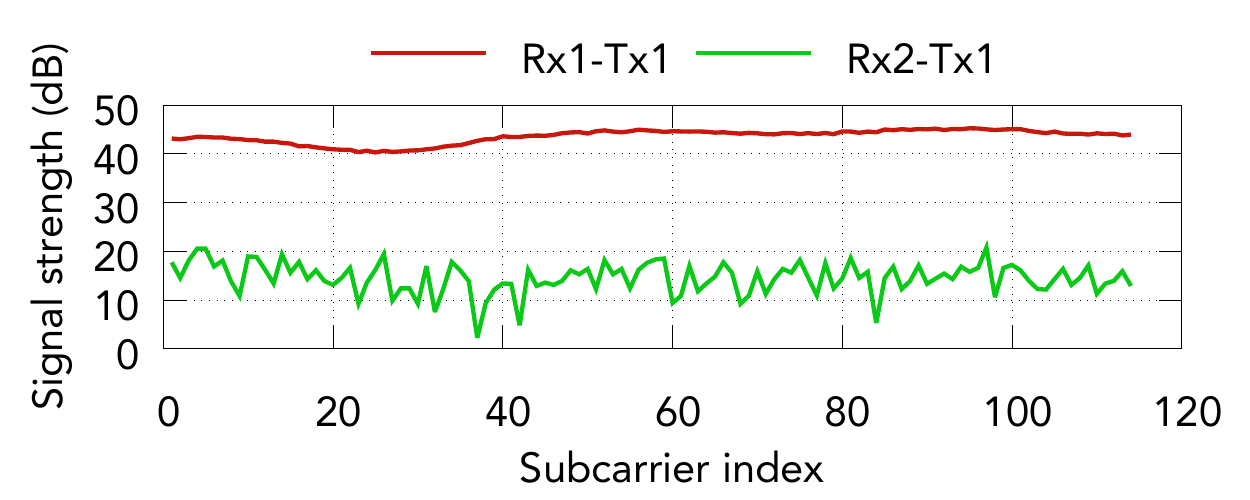}
	    \caption{{\bf Single-antenna baseline.} \textmd{Rx2 is neither connected to a Wi-Fi antenna nor the surface.}}  
\end{subfigure}
    \begin{subfigure}[b]{0.47\textwidth}
    \includegraphics[width=0.47\textwidth]{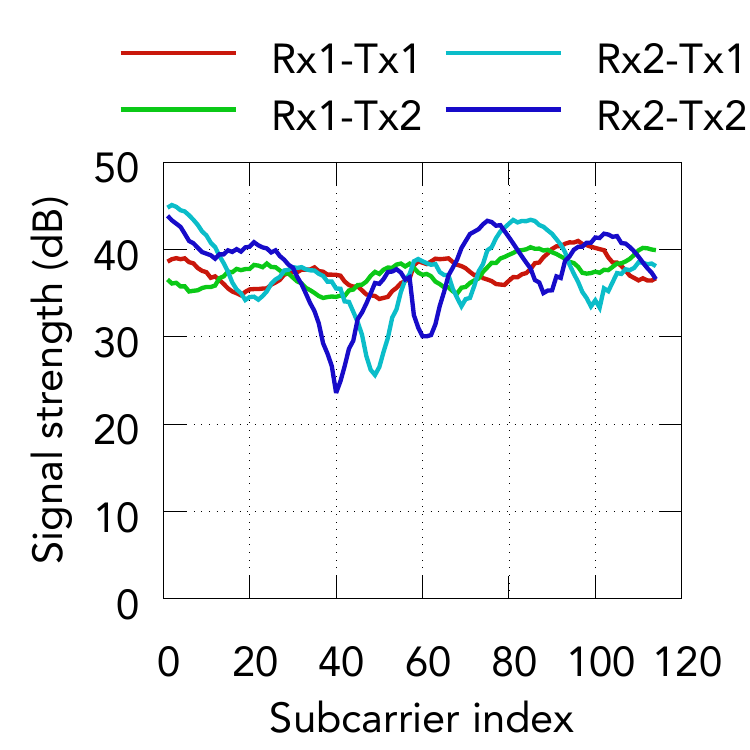}
    \includegraphics[width=0.47\textwidth]{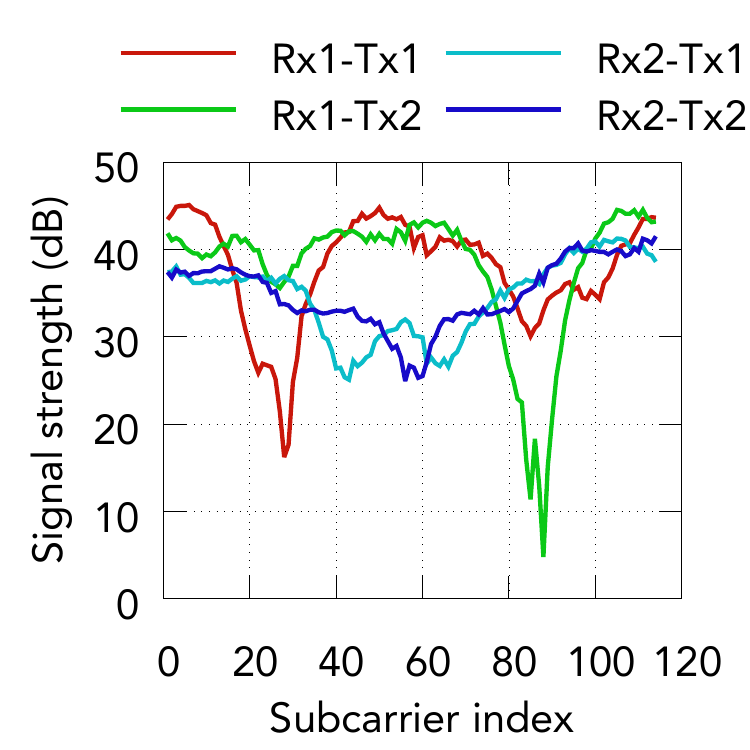}
    \caption{\textmd{MIMO over conductive spraypaint}}
    \end{subfigure}
  \caption[caption]{{\bf Channel state information.} \textmd{\red{Relative signal strength} with a surface MIMO setup using the conductive surface. Measured 1 feet (left) and 16 feet (right) from the transmitter on the surface.}}
\label{fig:csi}
\vspace{-0.1in}
\end{figure}

To understand the MIMO channel resulting from the conductive surface, we first measure the MIMO CSI when the transmitter and receiver are {\it not} in contact with the conductive surface. That is, they have a single Wi-Fi antenna and the second port has neither an antenna nor is in contact with the conductive surface. The transmitter and receiver are separated by 16~feet. Fig.~\ref{fig:csi}(a) shows that CSI for the signal propagation from the Wi-Fi antenna at the transmitter (Tx1) to the Wi-Fi antenna at the receiver (Rx1) and the unconnected port at the receiver (Rx2). As expected, the signal on the port without an antenna is significantly degraded.

\begin{figure*}[t]
\centering
    \begin{subfigure}[b]{0.44\textwidth}
    \includegraphics[width=\textwidth]{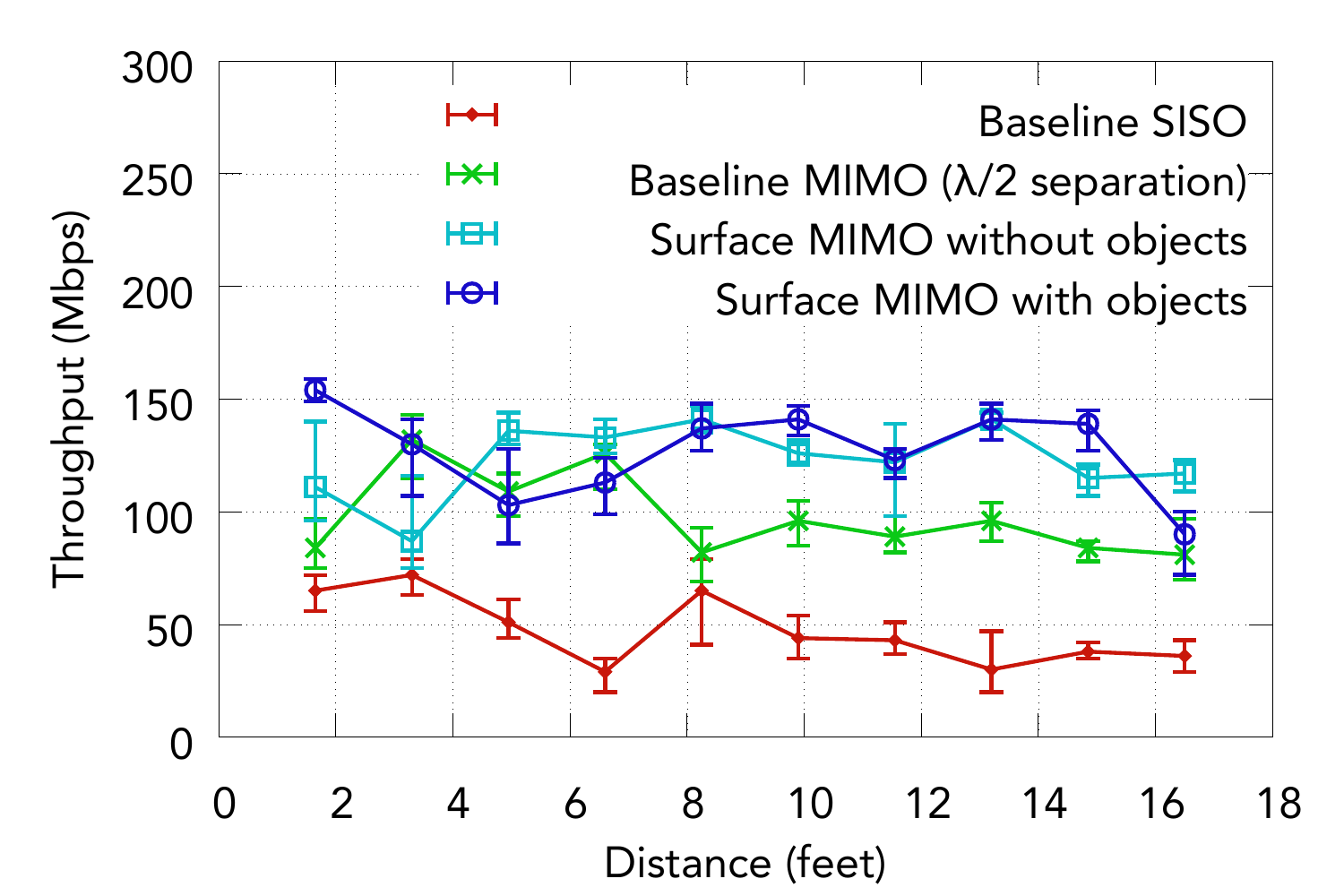}
    \caption{$2 \times 2$ spraypaint MIMO}
    \end{subfigure}
    \begin{subfigure}[b]{0.44\textwidth}
    \includegraphics[width=\textwidth]{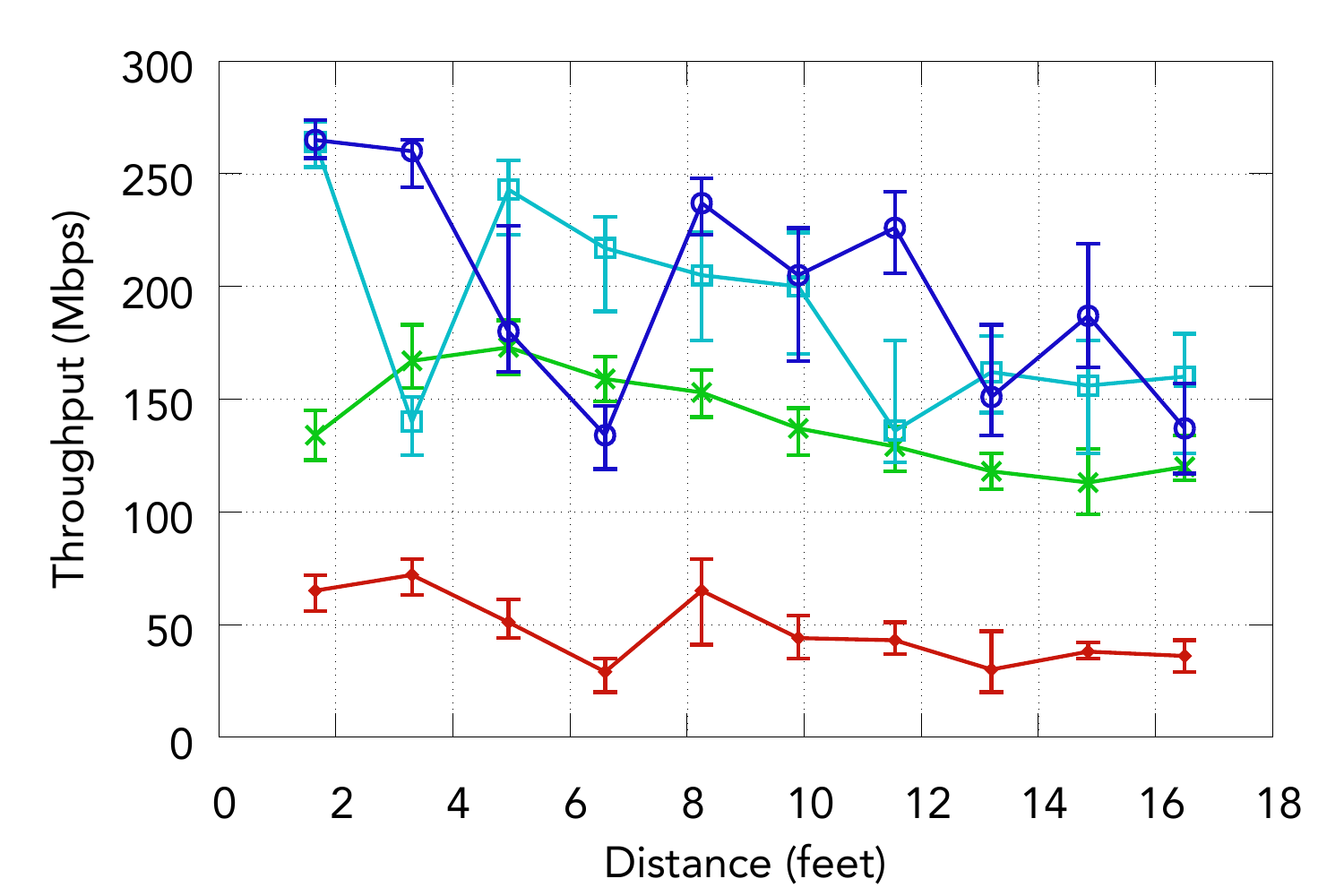}
    \caption{$3 \times 3$ spraypaint MIMO}
    \end{subfigure}
    
    \begin{subfigure}[b]{0.44\textwidth}
    \includegraphics[width=\textwidth]{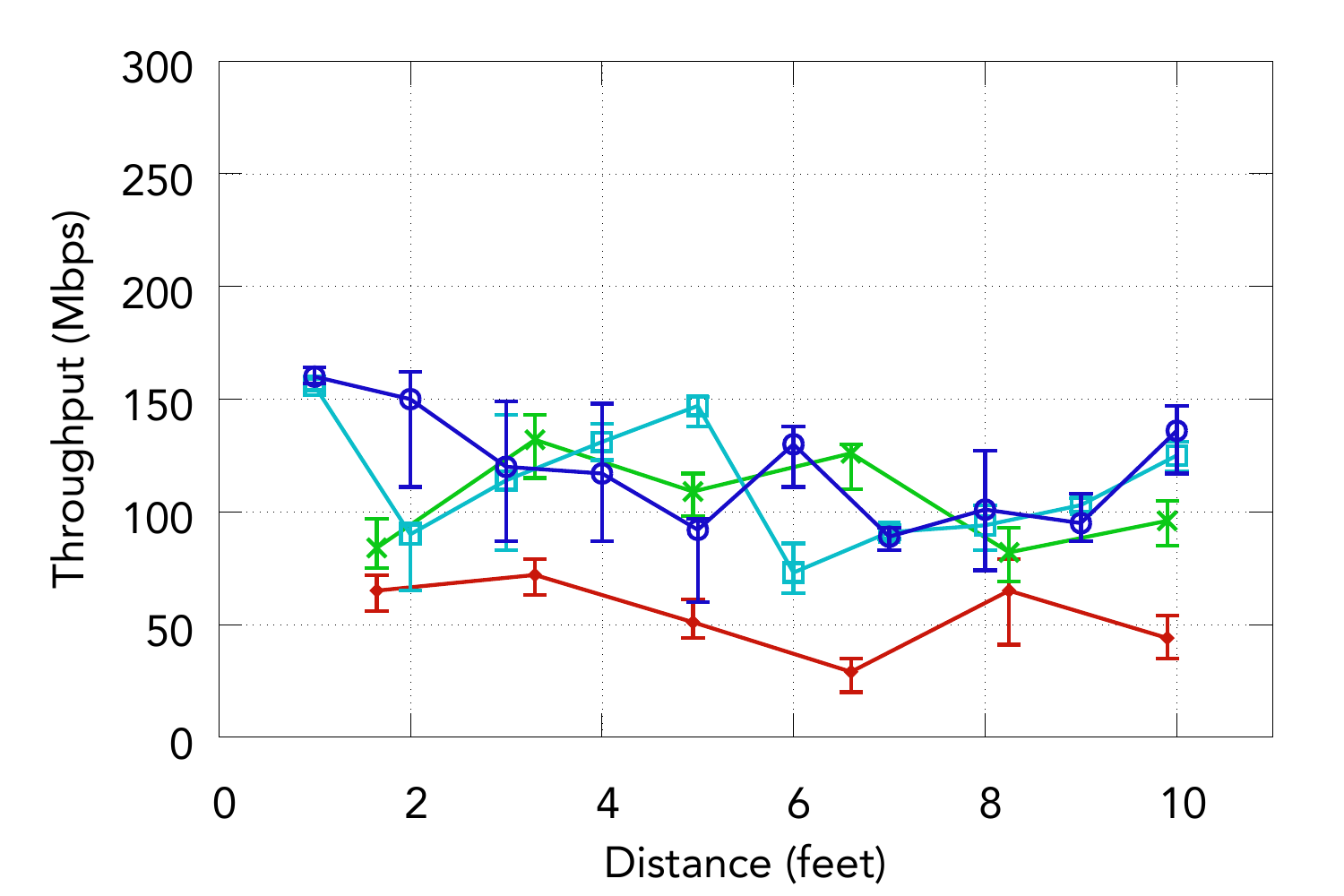}
    \caption{$2 \times 2$ cloth MIMO}
    \end{subfigure}
    \begin{subfigure}[b]{0.44\textwidth}
    \includegraphics[width=\textwidth]{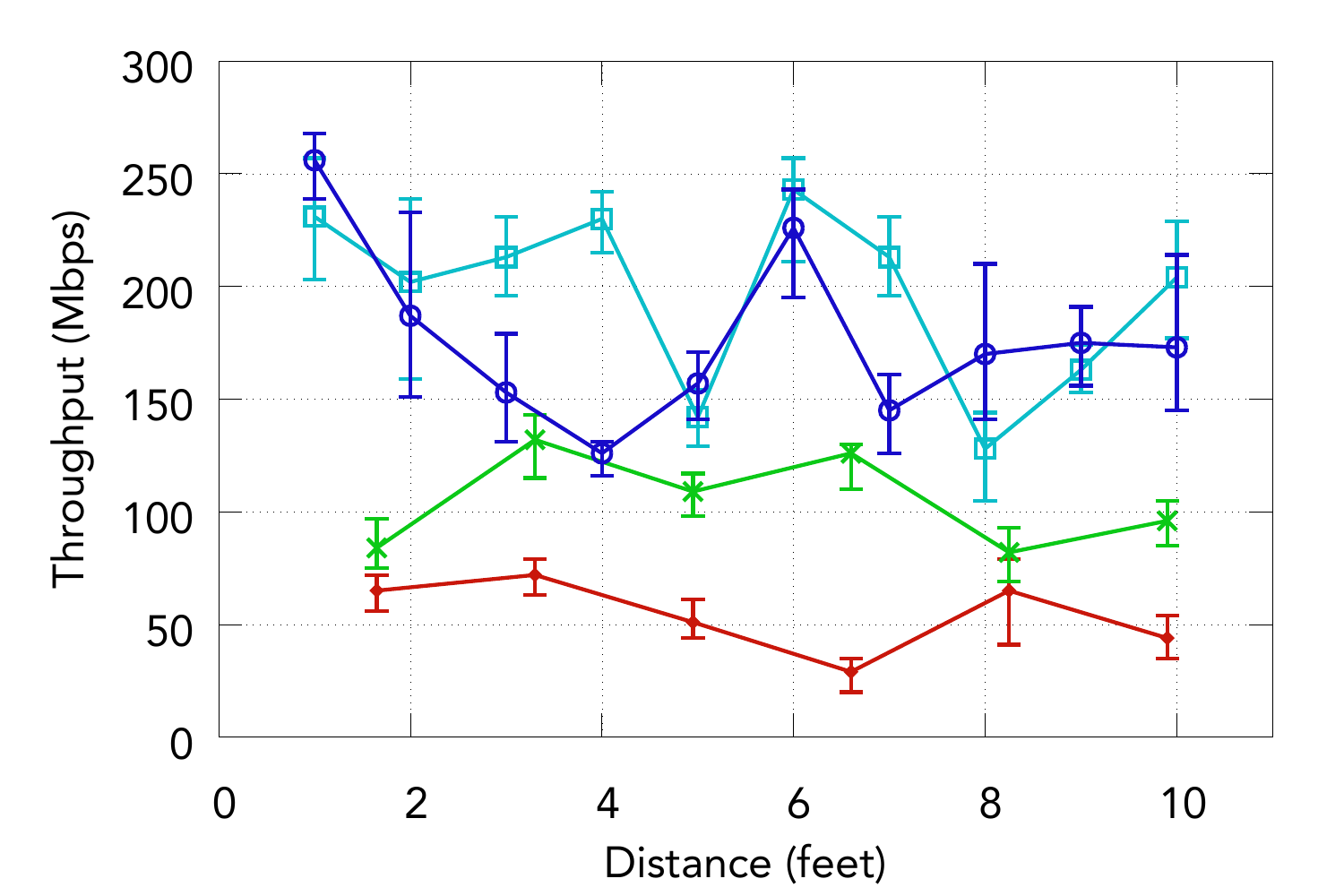}
    \caption{$3 \times 3$ cloth MIMO}
    \end{subfigure}
\vspace{-0.1in}
 \caption[caption]{{\bf Surface MIMO performance.} \textmd{Throughput results using a UDP connection.}}
\label{fig:udpmimo}
\vspace{-0.1in}
\end{figure*}

\begin{figure*}[t]
\centering
    \begin{subfigure}[b]{0.44\textwidth}
    \includegraphics[width=\textwidth]{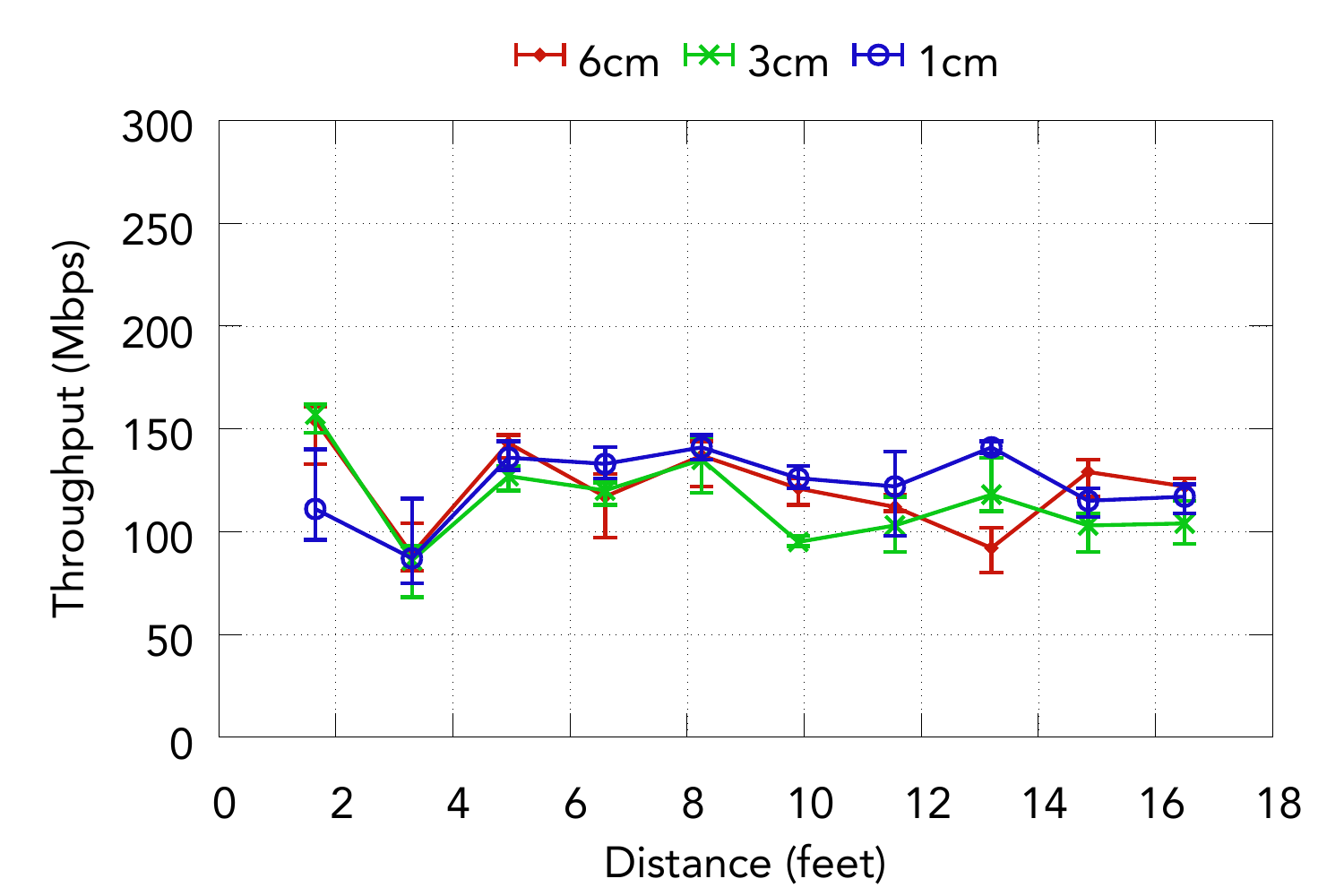}
    \caption{$2 \times 2$ spraypaint MIMO}
    \end{subfigure}
    \begin{subfigure}[b]{0.44\textwidth}
    \includegraphics[width=\textwidth]{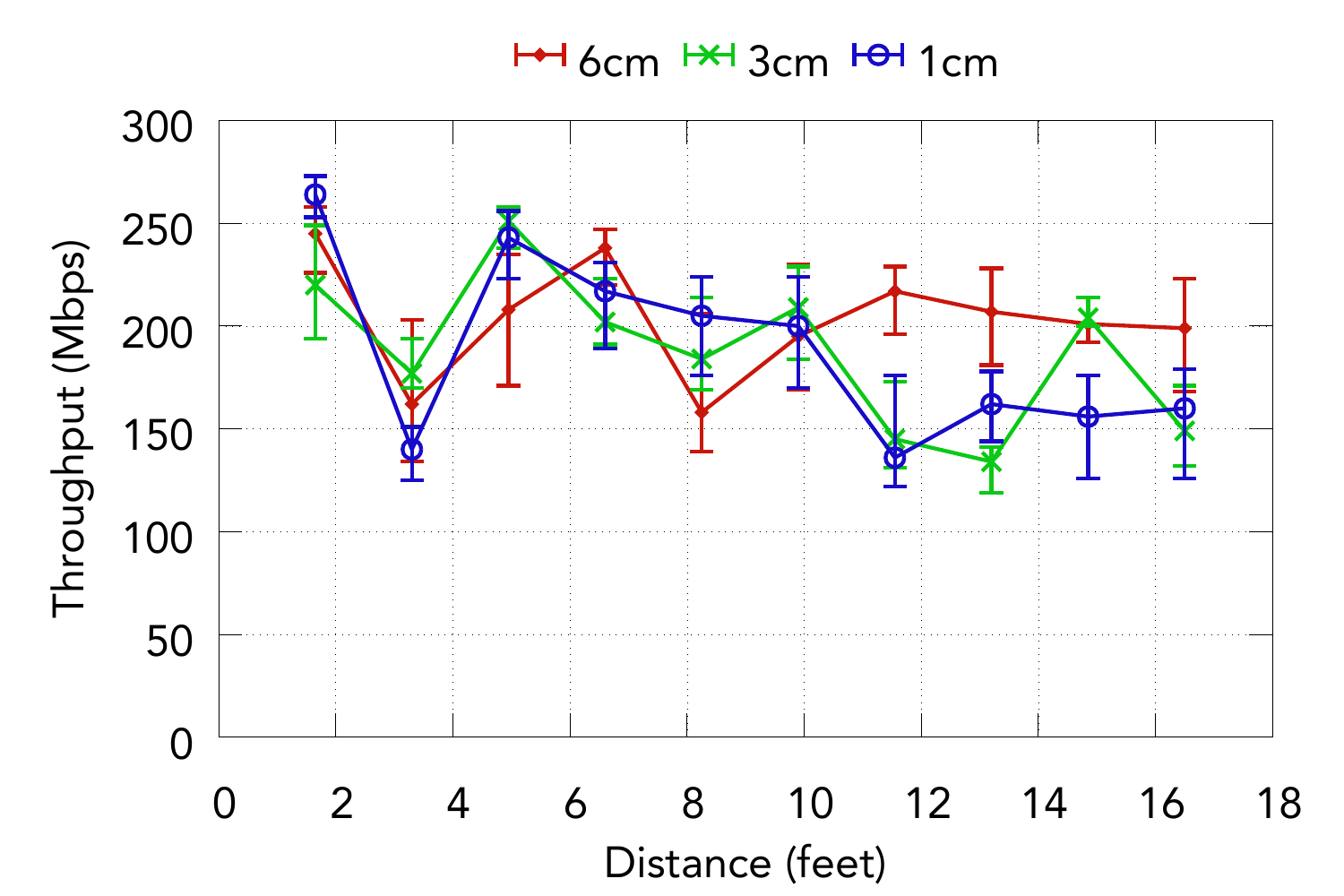}
    \caption{$3 \times 3$ spraypaint MIMO}
    \end{subfigure}
    
    \begin{subfigure}[b]{0.44\textwidth}
    \includegraphics[width=\textwidth]{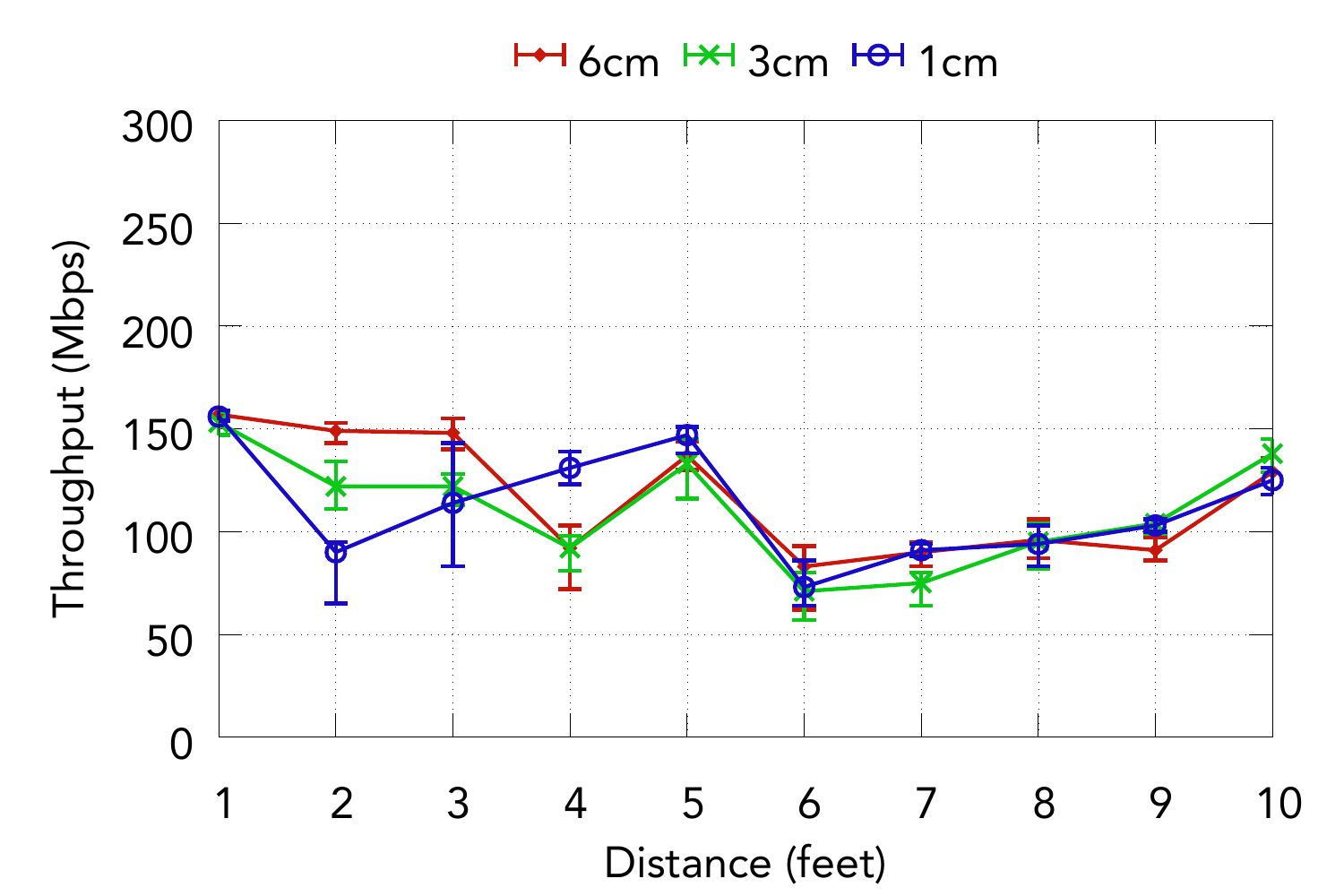}
    \caption{$2 \times 2$ cloth MIMO}
    \end{subfigure}
    \begin{subfigure}[b]{0.44\textwidth}
    \includegraphics[width=\textwidth]{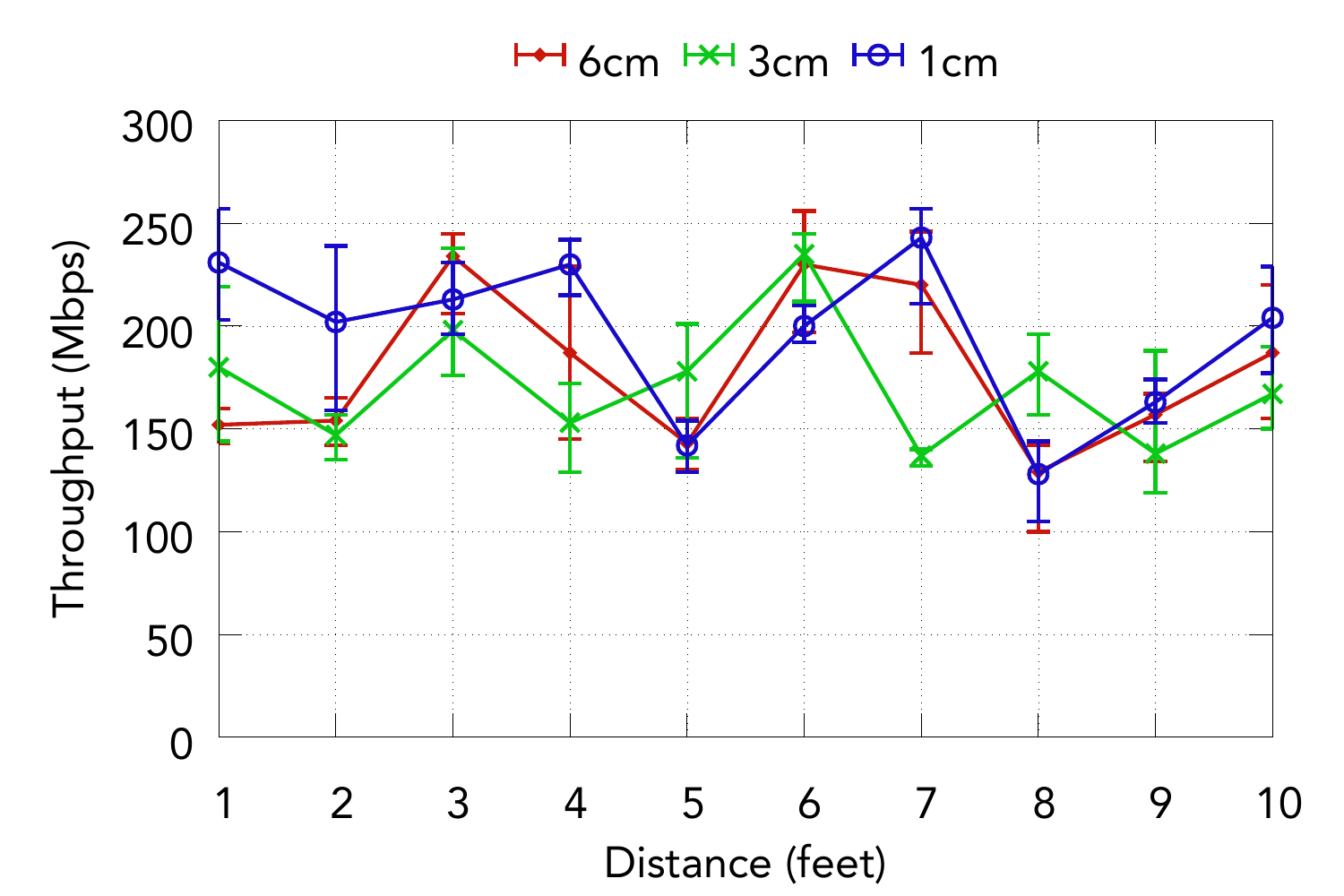}
    \caption{$3 \times 3$ cloth MIMO}
    \end{subfigure}
\vspace{-0.1in}
 \caption[caption]{{\bf Effect of separation.} \textmd{Throughput results when varying the separation between antennas and contacts.}}
\label{fig:sep}
\vspace{-0.1in}
 \end{figure*}
 
Next, we connect the second port on the transmitter and the receiver to the conductive surface. Fig.~\ref{fig:csi}(b) plots the CSI measurement across the four pair of antennas at two distances of 1 and 16~feet with the spraypaint as a conductive surface. The figures show that the signal strength between the two ports connected to the conductive surface is comparable to that between the Wi-Fi antennas. Further, the channels between the antennas are independent enough to support spatial multiplexing and thus a second stream.

The channel between the port connected to the surface at the transmitter (receiver) and the Wi-Fi antenna at the receiver (transmitter) is also surprisingly strong. This is because of near-field interaction between the conductive surface and the Wi-Fi antennas, which are next to each other.

As the distance increases, the MIMO channels experience more frequency diversity. This is expected given the multipath properties of the conductive surface that we reveal in~\xref{sec:surfacechar}. These results show that the MIMO channel created by the conductive surface could be used to support a second spatial stream for single antenna devices.

\subsubsection{End-to-end performance}
As before, we use Atheros  chips as our transmitter-receiver pair. We place the receiver at increasing distances from the transmitter while in contact with our conductive surface.  The transmitter and receiver {\bf do not share a common ground reference.}  We use two different baselines to compare with surface MIMO: 
\squishlist
\item {\it Baseline SISO.} Here we use one antenna at the transmitter and receiver for communication over the air.
\item {\it Baseline MIMO.} We use two antennas each at the transmitter and the receiver and use MIMO transmissions over the air.  The two antennas at the transmitter and receiver are separated by half a wavelength, which at 2.4~GHz is 6.25~cm.  
\squishend
We use 2.4~GHz PCB meandering antennas for in-air transmissions.  We measure the end-to-end UDP throughput for the above baselines while running an $iperf$ UDP test between the transmitter-receiver pair. We also collect a baseline throughput measurement for a $2 \times 2$ and $3 \times 3$ in-air baseline MIMO system using two and three PCB antennas each on the transmitter and receiver. For the baseline measurements, the PCB antennas were separated by half a wavelength, which at 2.4~GHz is 6.25~cm. 

We measured the throughput over distance of a $2 \times 2$ Surface MIMO system by attaching one PCB antenna and one contact to each transmitter and receiver pair. In this setup, the antenna and contact are only separated by 1~cm which is less than half a wavelength. Next we cluttered the surface with various objects including books and boxes and repeat the UDP throughput measurements. We repeat these throughput measurements for a $3 \times 3$ Surface MIMO system over conductive spraypaint and conductive cloth. \red{This setup uses one PCB antenna and two surface contacts. with a 1~cm separation.}

Fig.~\ref{fig:udpmimo} shows the results for the $2\times 2$ and $3\times 3$ surface MIMO scenarios over both the conductive surfaces.  The plots show the following.
Fig.~\ref{fig:udpmimo} shows the results for the $2\times 2$ and $3\times 3$ surface MIMO over both the conductive surfaces.  The figure shows the following.
\squishlist

\item In the case of spraypaint, a $2\times 2$ and $3\times 3$ Surface MIMO systems perform 2.6x and 3x better on average than the baseline SISO setup. This demonstrates the MIMO capability of our conductive surfaces.
\item The $2\times 2$ and $3\times 3$ surface MIMO perform 1.2x and 1.3x better on average than the baseline $2\times 2$ and $3\times 3$  over-the-air MIMO  setups, where the antennas were separated by half a wavelength.  This is for two reasons: 1) the surface acts like an antenna which provides a larger gain than the PCB antennas and 2) when the two devices are nearby, the multipath on the surface is stronger than multipath over air, allowing the Wi-Fi cards to obtain higher MIMO gains.
\item Additionally, placing objects (e.g., books and laptops) on the surface have no noticeable effect on the network throughput. This is because adding objects changes the multipath properties of the conductive surface but does not produce a significant relative SNR reduction at closeby distances on the air and does not significantly attenuate the signal propagation over the surfaces. Finally, there is significant variance across distance because of unevenness in the conductive material in the case of spraypaint and non-uniformly of the conductive material in the case of the cloth.

\squishend

\subsubsection{Effect of separation}
{Finally, we evaluate the effect that  separation between the antenna and the contact point has on UDP throughput. To do this, we first placed the PCB antenna and contact at the transmitter and receiver 6~cm (half a wavelength at 2.4~GHz) away from each other, then measured the end-to-end UDP throughput. We repeated this measurement for a 3~cm and 1~cm separation. Fig.~\ref{fig:sep} shows the throughput for different separations in a $2 \times 2$ and $3 \times 3$ Surface MIMO setup across conductive spraypaint and cloth. In traditional MIMO systems, antennas have to be separated by at least half a wavelength to achieve expected throughput gains. However, for surface MIMO, we find that there is no significant difference in throughput no matter what separation is used. This is because of two reasons a) the surface and air channel are independent so the separation between the PCB antenna and contacts have no effect b) the wavelength of the propagating wave on the surface is much shorter, so even when the surface contacts are separated by 1~cm, the system as a whole is still able to achieve high throughputs.}

\section{Gigabit communication}
\label{sec:gig}

Our goal here is to enable devices to communicate at Gbps link rates when they are in contact with the surface. Based on our characterization of conductive surfaces in~\xref{sec:char}, these surfaces can operate across a wide range of frequencies. One approach is to implement an Ethernet-like physical layer which can achieve Gbps speeds and uses frequencies up to 100s of MHz. The difference however is that unlike wired systems (e.g. Ethernet) where the copper cable is \red{shielded} to avoid unintentional radiations, our conductive surfaces are not \red{shielded}, and hence can radiate weak RF signals into the environment. Thus, if we were to implement an Ethernet-like system over conductive surfaces, it would be incompatible with FCC regulations since unlicensed radiations in the 1MHz to 900MHz range are restricted.

\begin{figure}[t]
\centering
\includegraphics[width=0.23\textwidth]{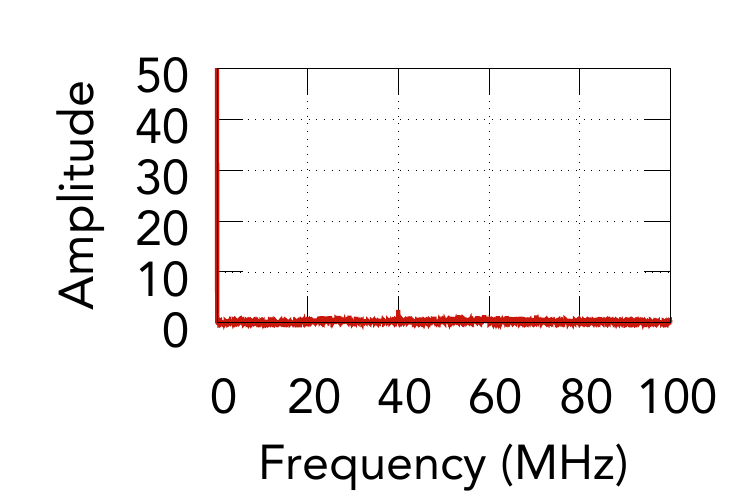}
\includegraphics[width=0.23\textwidth]{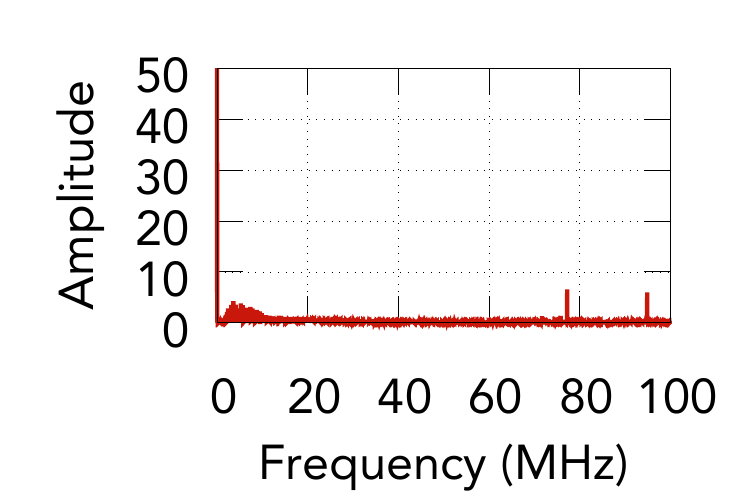}
\vspace{-0.1in}
\caption[caption]{{\bf Low frequency interference on conductive cloth.} \textmd{(Left) Anechoic chamber. (Right) Office room. We pick up weak FM radio transmissions in the office room.}}
\label{fig:interf}
\vspace{-0.1in}
\end{figure}

To see how much our conductive surfaces could interfere at these lower frequencies, we placed an SMA connector connected to a 1Gsps ADC onto a 1 by 1 feet piece of conductive cloth. The ADC provides us with raw digital samples on which we perform a FFT. Fig.~\ref{fig:interf} shows the frequency spectrum when the conductive surface is in an anechoic chamber and in an office setting. We see that even in the anechoic chamber, we see interference particularly in the $<$ 10MHz range which is a result of various machines and power lines creating interference at these frequencies. In the office setting, we also pick up frequencies in the FM bands demonstrating that our conductive surface can receive and hence by reciprocity radiate signals.

\begin{figure}[t]
\centering
\includegraphics[width=0.4\textwidth]{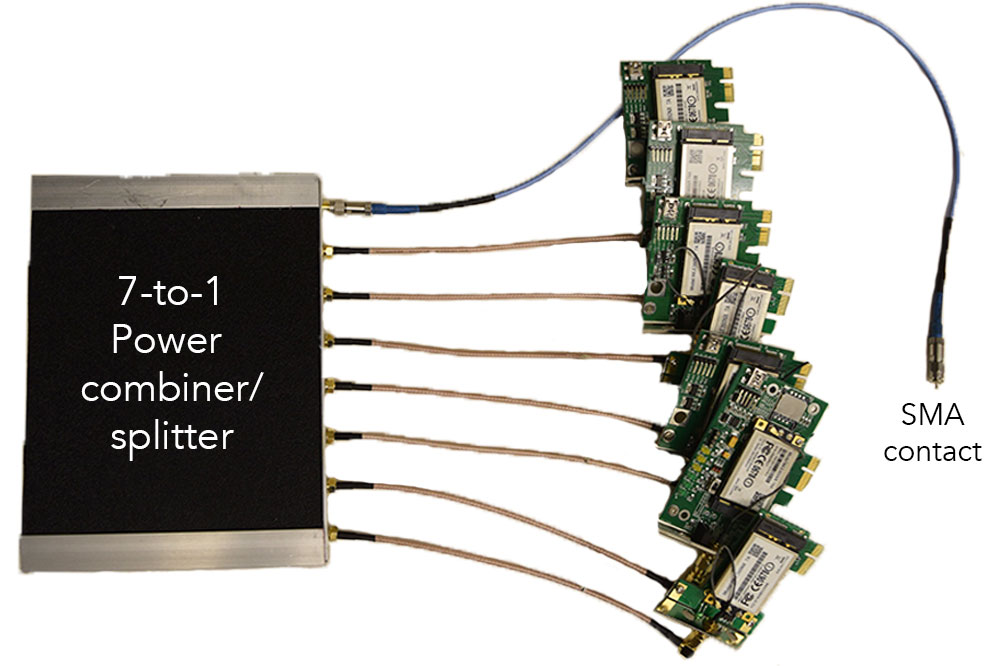}
\vskip -0.05in
\caption{{\bf Prototype Gigabit Hardware platform not built for size.} \textmd{Our platform using off-the-shelf Wi-Fi chipsets. All the cards are connected to the surface using a single contact. Note that in practice one would design an ASIC that can go into small devices.}}
\label{fig:proto}
\vspace{-0.1in}
\end{figure}

We build a hardware platform shown in Fig.~\ref{fig:proto} that enables us to use the bandwidth across all the ISM bands over the conductive surfaces. Instead of using expensive software radios that cannot currently handle such a wide bandwidth, we build our platform using off-the-shelf Wi-Fi cards. Specifically, 802.11ac Wi-Fi chipsets can support wide bandwidths using OFDM and can tolerate multipath delays of up to 800~ns. Our hardware combines the outputs of seven Wi-Fi chips using a ZN8PD1-63W+~\cite{combiner} seven-to-one power combiner and splitter to create a single tiny contact that can be placed in contact with our conductive surface. We use 802.11n and 802.11ac Wi-Fi chips to support various combinations of 40~MHz and 20~MHz channels. To generate a 900~MHz signal using commodity Wi-Fi cards we use the ZX05-63LH+~\cite{converter} wideband frequency mixer to downconvert a 2.4~GHz Wi-Fi signal to the 900~MHz ISM band. Specifically, we set our Wi-Fi cards to a center frequency of 2.412~GHz (Channel 1) and downconvert this to a center frequency of 915~MHz. As the 900~MHz ISM band spans 26~MHz, we set the bandwidth of our 2.4~GHz signal to 20~MHz. At the receiver, we use the same frequency mixer to upconvert the 900~MHz to a 2.4~GHz signal that can be decoded by Wi-Fi cards. The process of downconverting and upconverting results in a 6~dBm loss.

\begin{figure}[t]
  \centering
\begin{subfigure}[b]{0.47\textwidth}
    \includegraphics[width=\textwidth]{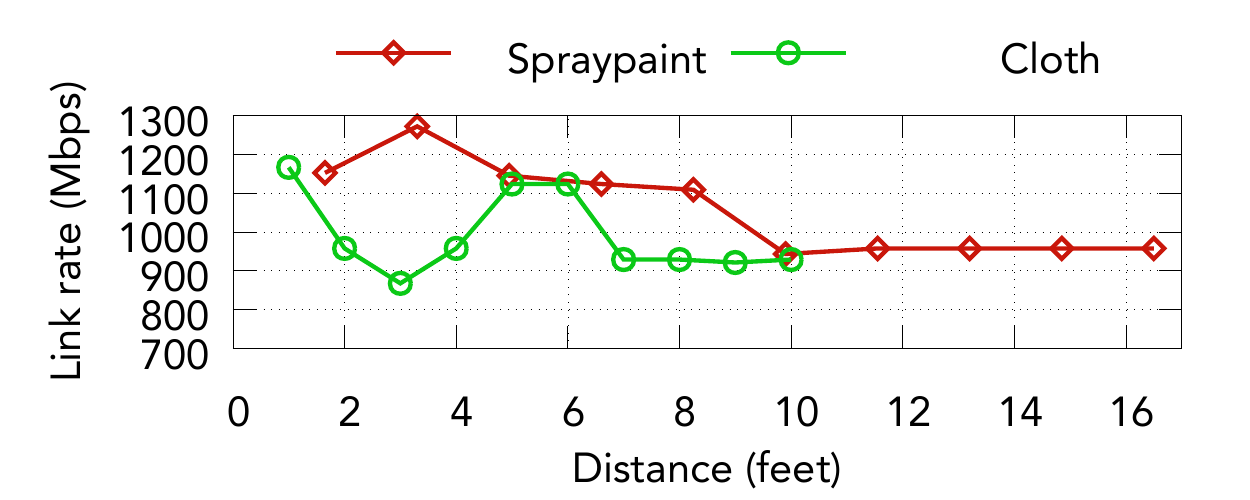}
    \caption{\textmd{Scenario 1}}
    \end{subfigure}
\begin{subfigure}[b]{0.47\textwidth}
    \includegraphics[width=\textwidth]{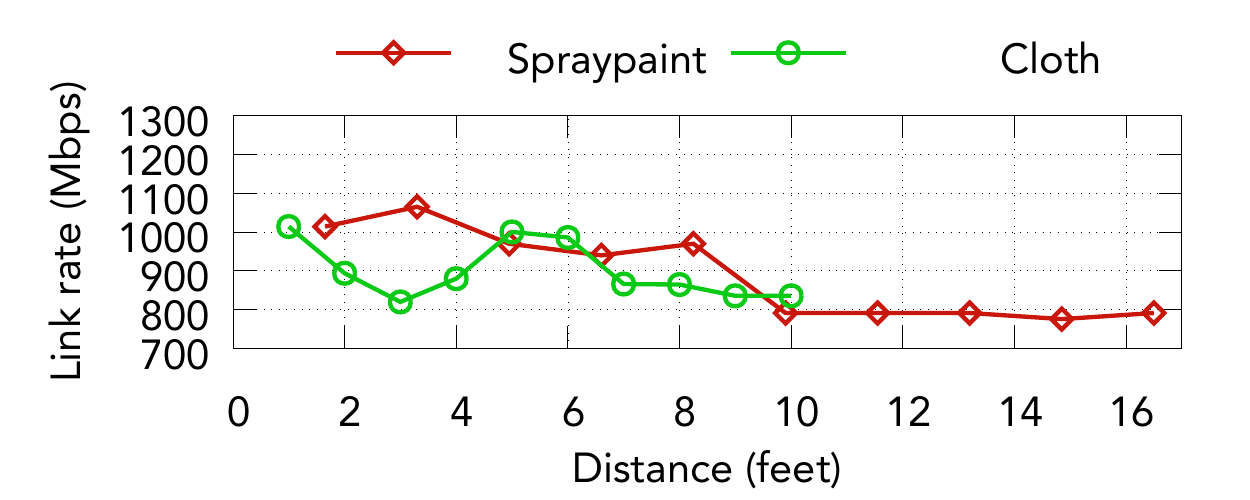}
    \caption{\textmd{Scenario 2}}
    \end{subfigure}
\vspace{-0.1in}
    \caption{{\bf Channel capacities} \textmd{Both the conductive surfaces support link rates between 776 Mbps and 1.27 Gbps across the whole length of the surface.}}
\label{fig:cap}
\vspace{-0.1in}
\end{figure}

{\it Capacity Evaluation.} We set one of our devices to be a Wi-Fi AP and the other to act as the client. We fix the location of the AP and increase its distance from the client. Each of the Wi-Fi chips uses Atheros’ ath10k firmware that allows us to set various properties including the frequency band, center frequency as well as OFDM guard times. We evaluate two different configurations for our seven Wi-Fi cards, \red{the first makes use of DFS channels and the second does not:}
\squishlist
\item {\it Scenario 1:} We set six of the Wi-Fi chipsets to a bandwidth of 40~MHz and the seventh to 20~MHz \red{channel} of the  5~GHz ISM band that does not interfere with weather radar between channels 116 to 136 \red{for a total bandwidth of 260~MHz}. We use two 40~MHz DFS (dynamic frequency spectrum) channels on 5~GHz. Devices intending to use DFS channels are required to check for the presence of any radar transmissions on these frequencies before using them.

\item {\it Scenario 2:} We set four of the Wi-Fi chipsets to non overlapping 5~GHz channels each with a bandwidth of 40~MHz. We do not use any of the DFS channels. We then set two of the Wi-Fi chipsets to 2.4~GHz: one with a bandwidth of 40~MHz and the other at the non-overlapping 20~MHz channel. The seventh Wi-Fi chipset is set to transmit at 900~MHz with a 20~MHz bandwidth. This results in a total bandwidth of 240~MHz. 
\squishend

\vskip 0.05in\noindent{\it Our design.} Our approach is to use the transmissions in the 900~MHz, 2.4 and 5~GHz ISM band.

Each of the Wi-Fi chipsets is set to an adjacent Wi-Fi channel and independently sends data to its counterpart at the receiver.  We allow each of the Wi-Fi chipsets to automatically pick their Wi-Fi guard durations as well as the PHY bit rate. We compute the capacity supported by our hardware as the sum of the bit rates on each of the Wi-Fi cards at which the received error rate is less than 1\%.  Fig.~\ref{fig:cap} plots the capacity for each of these scenarios. The figure shows that both the conductive surfaces support high link rates between 776~Mbps and 1.27~Gbps across the whole length of the surface. We note that the actual throughput achieved by the devices, as is the case with Wi-Fi, will be determined by the number of other devices on the network.

\section{Microbenchmarks}
We benchmark communication over conductive surface.
\begin{figure}[t!]
\centering
\includegraphics[width=0.4\textwidth]{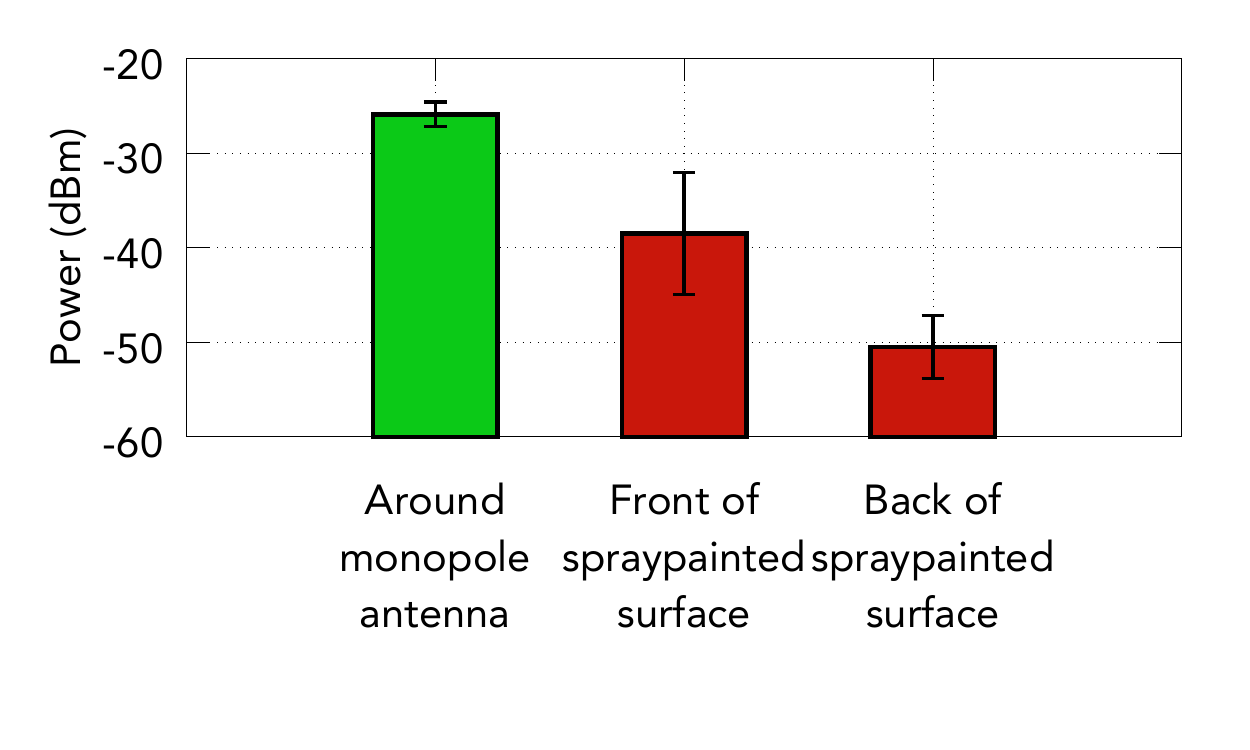}
\vspace{-0.4in}
\caption[caption]{{\bf In-air interference.} \textmd{Power is distributed symmetrically around a monopole antenna. However, power radiation is concentrated only in the front of the spraypainted surface. Further the power is 13 to 25~dBm worse on the air with our conductive surfaces. This translates to a reduction of in-air interference.}}
\label{fig:antenna}
\vspace{-0.1in}
\end{figure}

{\bf In-air interference.} The surface acts as an antenna and absorbs RF signals over the air as well. Hence, over-the-air transmissions in the same frequency band (\textit{e.g.}, WiFi signals) introduce interference on surface communications. We measure the amount of signal that leaks into the air. While the results are similar across the ISM band frequencies, we plot the results at 2.4~GHz. We compare two cases: 1) when the transmitter and receiver both use 0~dBi antennas and 2) when the transmitter is placed in contact with the conductive surface and the receiver uses a 0~dBi  antenna. In the case when both the transmitter and receiver use 0~dBi antennas, the propagation is symmetric in all directions. However in the case of the conductive surface, the propagation is different on the front and the back of the surface. 

Thus, Fig.~\ref{fig:antenna} shows the average signal strength for three scenarios: 1) regular Tx-Rx pair which both use 0~dBi antennas, 2) Tx is placed in contact with the conductive surface and Rx is a 0~dBi antenna placed in front of the conductive surface and 3) Tx is placed in contact with the conductive surface and Rx is a 0~dBi antenna placed in front of the conductive surface. The plot shows that the signals on the front of the conductive surface are attenuated by 13~dB in comparison to when using a normal antenna. The attenuation is even better at 25~dB on the back of the conductive surface. This demonstrates that when two devices share a surface, it is better for them to use surface communication since they create less interference in the air when compared to using antennas at the transmitter and the receiver. Thus, surface communication can be used to reduce the amount of interference on the wireless medium.

\begin{figure}[t]
\centering
\begin{subfigure}[b]{0.23\textwidth}
\includegraphics[width=\textwidth]{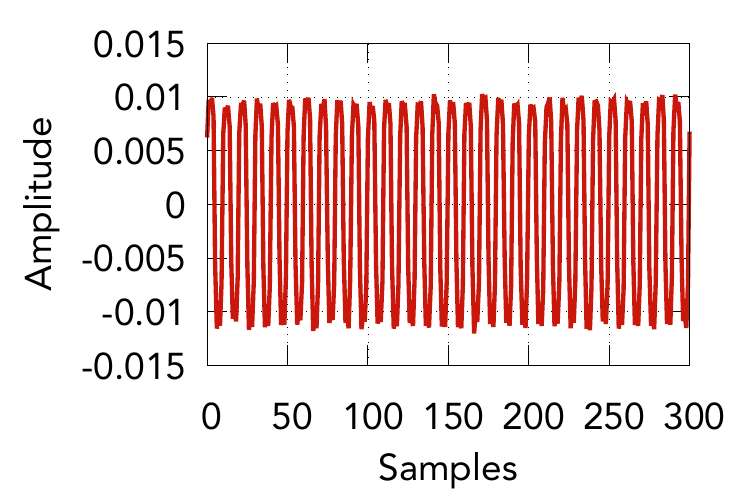}
\caption{\textmd{\red{Common ground}}}
\end{subfigure}
\begin{subfigure}[b]{0.23\textwidth}
\includegraphics[width=\textwidth]{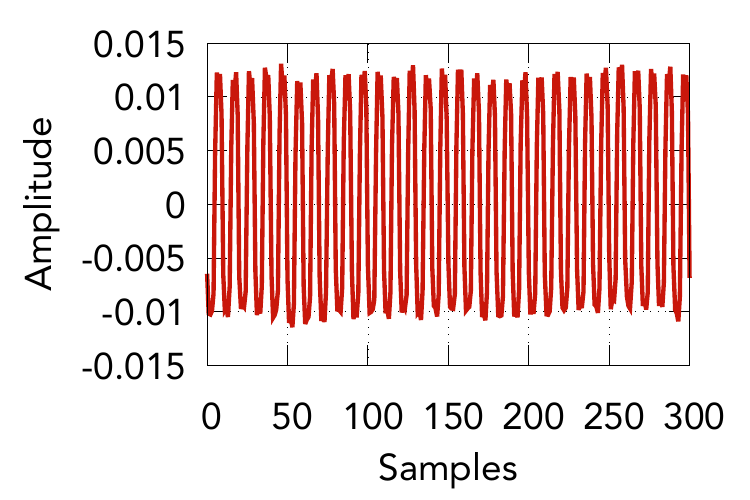}
\caption{\textmd{\red{Separated ground}}}
\end{subfigure}
\caption{\red{Effect of common and separated grounds on a 2.4~GHz signal transmitted on a spraypainted sheet.}}
\label{fig:gnd-sp}
\vspace{-0.1in}
\end{figure}

\red{
{\bf Effects of isolated ground.} A cable has at least two wires, one with the signal and the other providing a reference ground. Since our conductive surfaces are a single medium, we do not have a common ground, in particular, when the devices are battery-powered. To measure the effect of the lack of a common ground, we compare two scenarios: battery-powered transmitter with no common ground and  when we plug-in the transmitter and receiver into a wall outlet. Fig.~\ref{fig:gnd-sp} shows  2.4~GHz time-domain signals for our spray-painted surface. The plots shows that at 2.4~GHz, there is little change to the signal even when the transmitter and receiver do not share a common ground. 
}

\begin{figure}[t]
\centering
\begin{subfigure}[b]{0.47\textwidth}
\includegraphics[width=\textwidth]{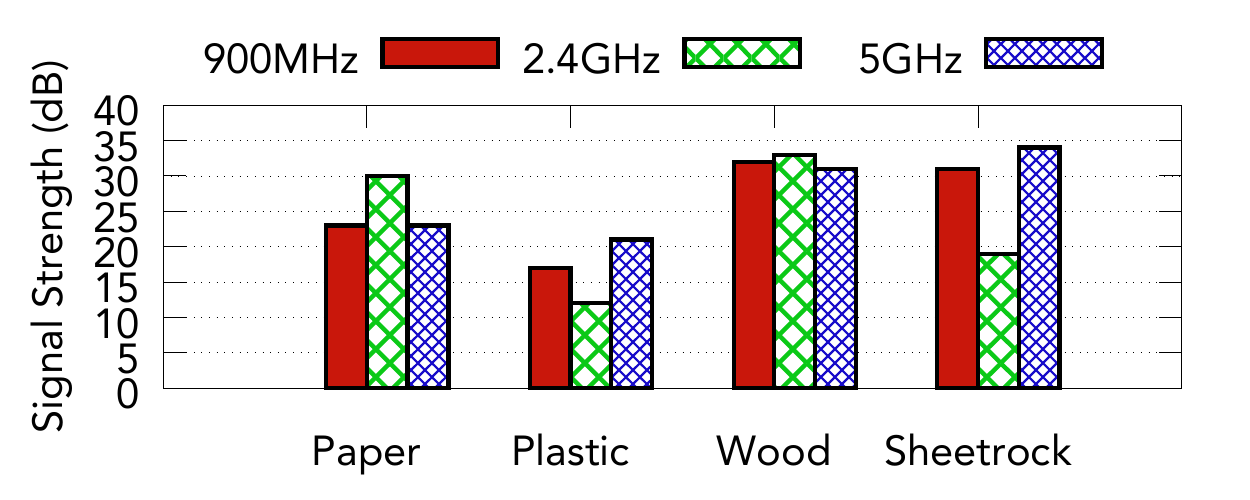}
\caption{\textmd{Conductive spraypaint}}
\end{subfigure}
\begin{subfigure}[b]{0.47\textwidth}
\includegraphics[width=\textwidth]{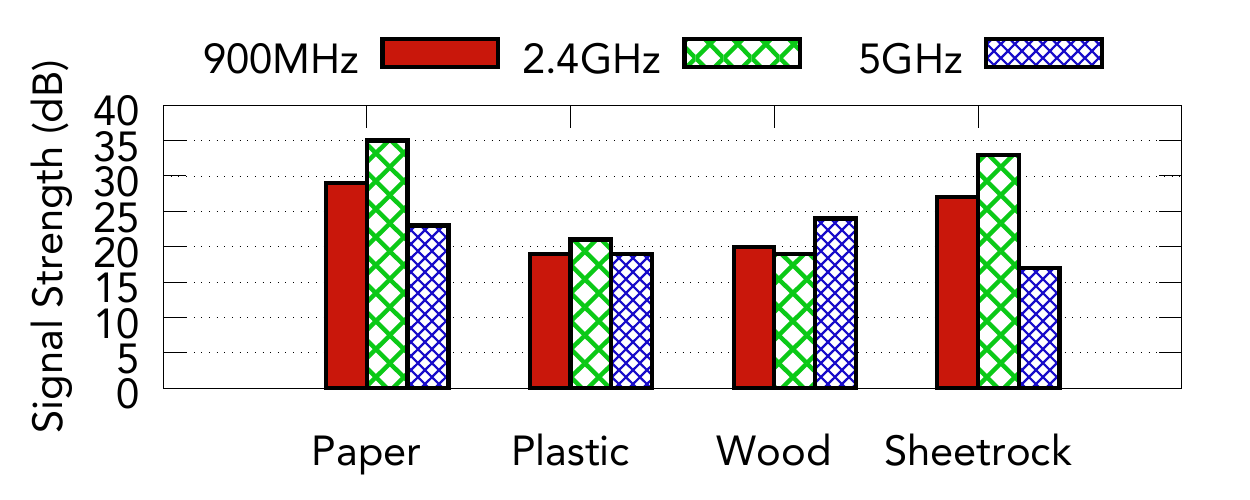}
\caption{\textmd{Conductive cloth}}
\end{subfigure}
\vspace{-0.1in}
\caption[caption]{Signal strength over different substrates. \textmd{\red{The transmitter and receiver are separated by a foot.}}}
\label{fig:sub}
\vspace{-0.1in}
\end{figure}

{\bf Effect of surface material.}
We measure the effect that different surface materials have on communication. To do this, we spraypaint different materials: paper, plastic, wood and sheetrock and send a \red{continuous wave signal at 900~MHz, 2.4~GHz and 5~GHz} through the surface. \red{We separate the transmitter and receiver by a distance of 1 feet} and measure the resultant SNR. We compute the SNR by comparing the signal power with that of noise. We do this same test with conductive cloth by placing the cloth over these surfaces. Fig.~\ref{fig:sub} shows the SNR for diferent materials. The highest SNR across all substrates for both conductive surfaces was 35~dB, while the lowest SNR was 12~dB. As seen in these measurements, the SNR can vary between material and frequency. However, we found in our experiments that the most important factor in achieving a high SNR is achieving good contact with the SMA connector and the conductive surface. As the conductive surfaces are inherently not uniform in nature, certain sections of the surface yield a higher SNR than others. Our results show that for the case of conductive spraypaint, it is easier to make good contact for certain substrates over others. \red{In our experiments we found that there were two requirements for good contact: a) a uniform coating or covering of conductive material over the substrate and b) firm contact between the device and substrate such that the two are visibly touching.} For example, the SNR for conductive spraypaint over wood and sheetrock were higher than when plastic was used as a substrate. In the case of conductive cloth, paper and sheetrock served as better contacts than plastic and wood. However, the SNR is still greater than 12~dB across all the materials.

\begin{figure}[t]
  \centering
    \begin{subfigure}[b]{0.23\textwidth}
    \includegraphics[width=\textwidth]{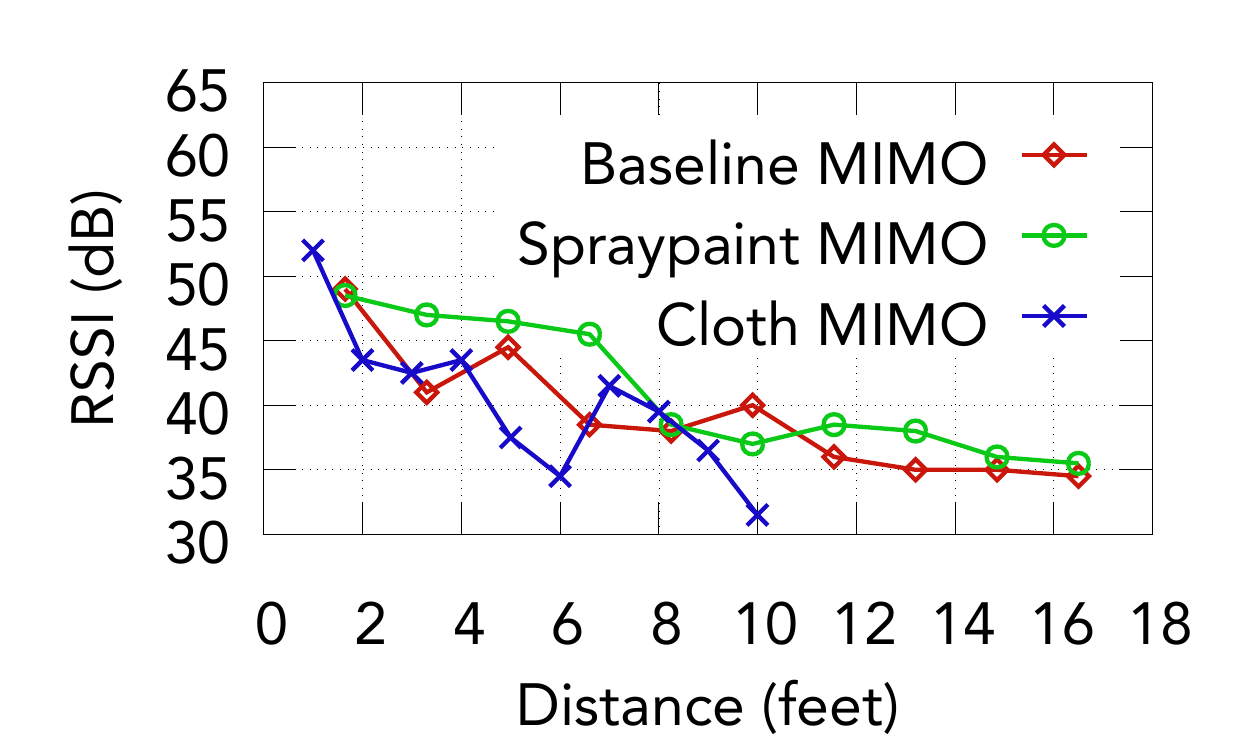}
    \caption{\red{2x2 MIMO}}
    \end{subfigure}
    \begin{subfigure}[b]{0.23\textwidth}
    \includegraphics[width=\textwidth]{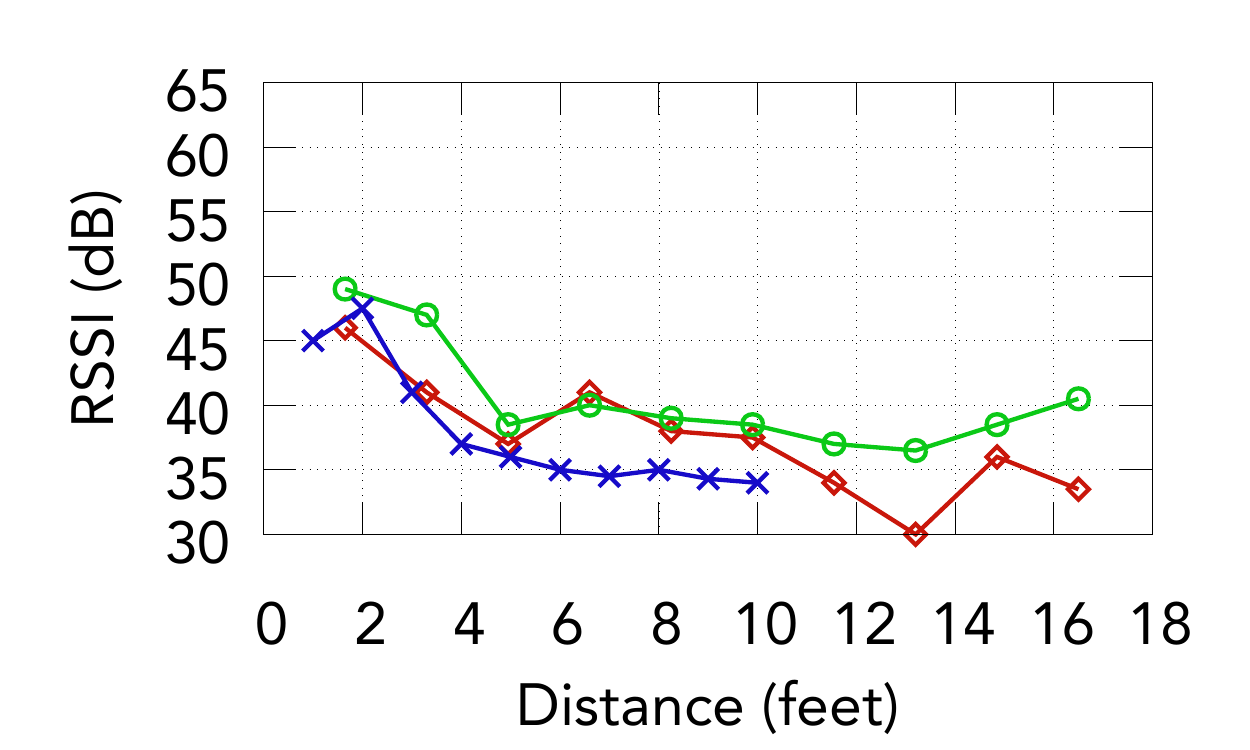}
    \caption{\red{3x3 MIMO}}
    \end{subfigure}
    \caption{\red{RSSI over distance. \textmd{}}}
\label{fig:rss}
\vspace{-0.1in}
\end{figure}

\begin{figure}[t]
  \centering
    \begin{subfigure}[b]{0.23\textwidth}
    \includegraphics[width=\textwidth]{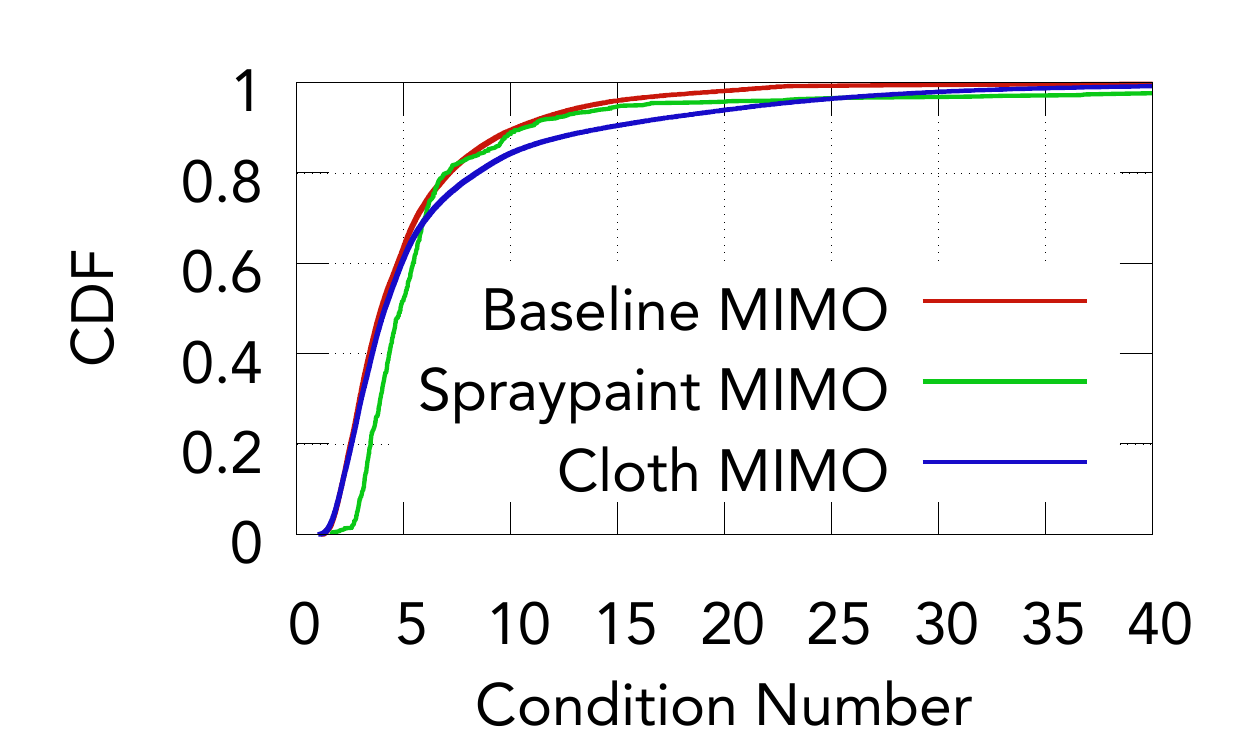}
    \caption{\red{2x2 MIMO}}
    \end{subfigure}
    \begin{subfigure}[b]{0.23\textwidth}
    \includegraphics[width=\textwidth]{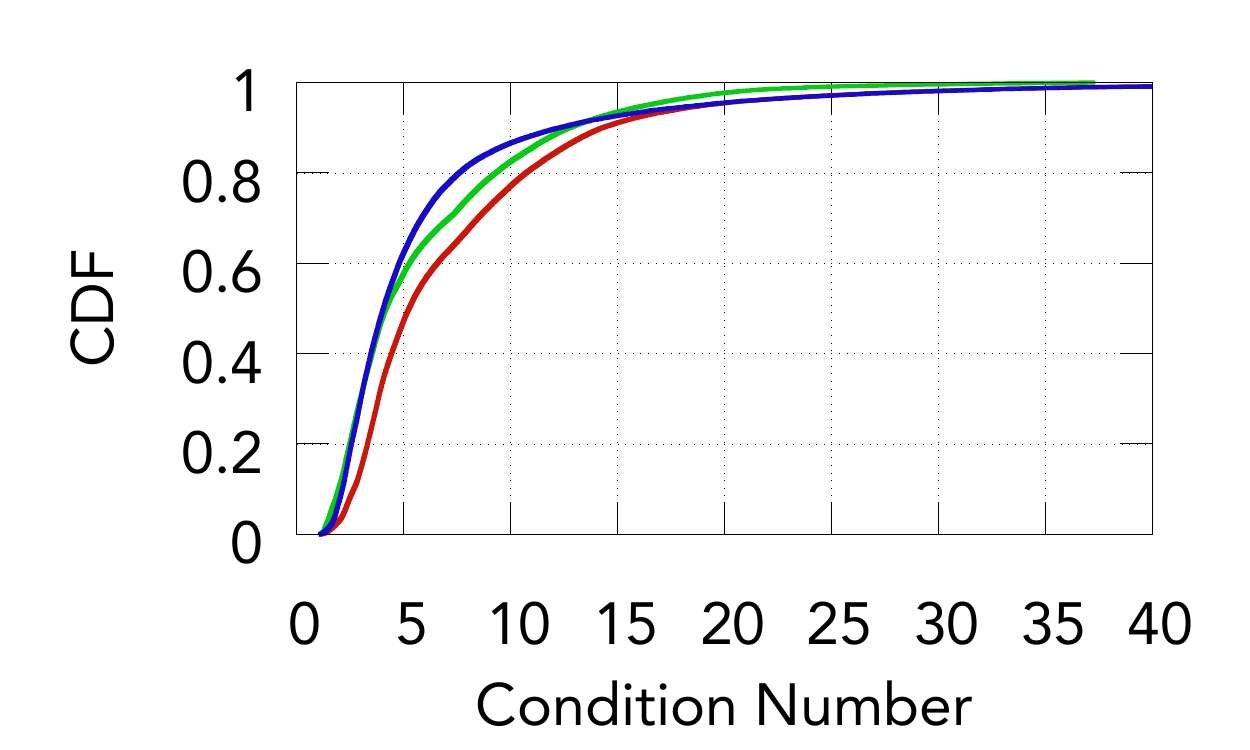}
    \caption{\red{3x3 MIMO}}
    \end{subfigure}
    \vskip -0.1in
    \caption{\red{Condition number of Surface MIMO. \textmd{}}}
\label{fig:eigen}
\vspace{-0.1in}
\end{figure}

\red{{\bf MIMO Channel analysis.} We measure the  channel's RSSI and condition number over distance. We do this for the baseline MIMO setup of PCB antennas with a half-wavelength separation and with Surface MIMO on conductive spraypaint and cloth. Fig.~\ref{fig:rss} shows the drop in average RSSI across spatial streams over distance for 2x2 and 3x3 MIMO configurations. Across all configurations, we observe an average decrease in RSSI of 13~dB across the entire measured distance. We note that the RSSI for conductive spraypaint is on average higher than the baseline by 2.3~dB in the 2x2 MIMO case and 3.2~dB in the 3x3 MIMO case. The RSSI for conductive cloth is on average slightly higher than the baseline by 0.4~dB and 0.7~dB in the 2x2 and 3x3 MIMO configurations respectively. Fig.~\ref{fig:eigen} shows the condition number of the channel matrix calculated across all OFDM subcarriers, spatial streams and packets. A condition number has to be small in low to medium SNR regimes but can be larger in high SNR regimes for MIMO communication~\cite{tse2005fundamentals}. We find that in all configurations including the baseline case there is no noticeable trend in the condition number over the distances measured. One reason for this could be that the measured distances were all relatively short. We also find that there were no statistically significant differences in the condition number between the baseline MIMO case and the surface MIMO cases. This evaluation indicates that for our surface MIMO channel can be used for communication in 2x2 and 3x3 use-cases. A more extensive evaluation with larger MIMO systems with four or more antennas is left for future work.}

\red{{\bf Sharing the surface.} The design in~\xref{sec:gig} focuses on achieving a high link rate between a single transmitter and receiver pair. However the high capacity supported by these surfaces can also be shared across multiple devices on the same surface. At a high level each device performs carrier sense before transmitting, in a similar manner to over the air Wi-Fi.}

\begin{figure}[t]
  \centering
    \begin{subfigure}[b]{0.23\textwidth}
    \includegraphics[width=\textwidth]{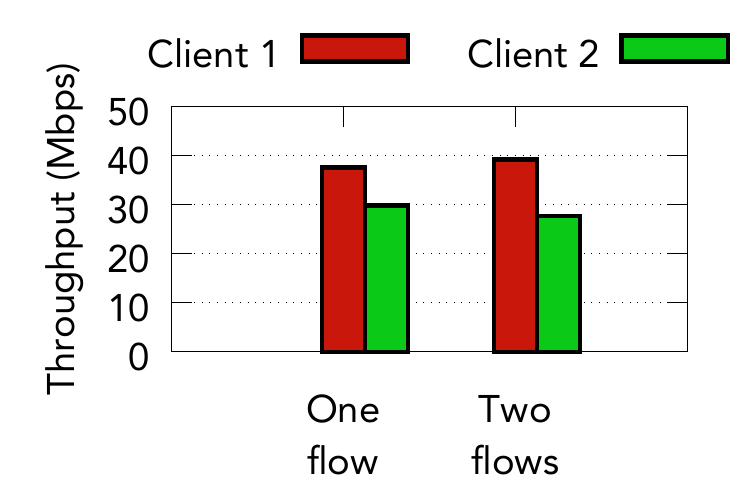}
    \caption{\textmd{\red{Different channels}}}
    \end{subfigure}
    \begin{subfigure}[b]{0.23\textwidth}
    \includegraphics[width=\textwidth]{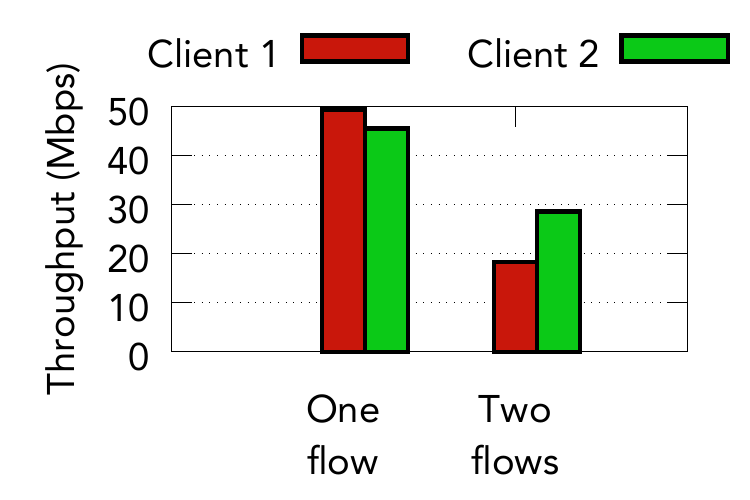}
    \caption{\textmd{\red{Same channel}}}
    \end{subfigure}
\vspace{-0.1in}
  \caption[caption]{\red{{\bf Sharing the surface.} \textmd{Both the clients are in contact with the surface and use carrier sense to share the conductive surface. The first plot is when the two clients are on channel 1 and 6 respectively and the second plot is when they share the same Wi-Fi channel 1.}}}
\label{fig:network}
\vspace{-0.1in}
\end{figure}

\red{To test how well such a design works we run networking experiments with two Wi-Fi transmitter-receiver pairs on the same surface. All the Wi-Fi devices do not have an antenna and instead are in contact with the conductive surface. Specifically, transmitters are set as clients and the receivers are set as APs. The client runs UDP flows using $iperf$ and computes the average throughput. The Wi-Fi clients are set to use 20~MHz of bandwidth. We evaluate this setup when the clients are on separate 2.4~GHz Wi-Fi channels (1 and 6)  and when they are on the same channel (1). The APs are placed next to each other at one end of our conductive spraypaint testbed and the two clients are placed two feet away from the APs. We run experiments in three scenario: when client 1 alone transmits, when client 2 alone transmits and when both client 1 and 2 transmit.}

\red{Fig.~\ref{fig:network} shows the results for both these scenarios. Fig.~\ref{fig:network}(a) shows that when the two clients are on different channels, running concurrent transmissions does not create interference and the clients get a similar throughput to when they are transmitting by themselves. This is expected because they use non-overlapping channels. We note however that the throughput achieved is around 30--40 Mbps because the wireless channel is being shared not just between the two clients in contact with the surface but also the Wi-Fi nodes in the environment. Specifically, since the conductive surface picks up strong Wi-Fi signals in the environment, carrier sense avoids collisions with conventional over-the-air Wi-Fi. Fig.~\ref{fig:network}(b) shows the throughput when the two nodes are on the same channel and show that both the clients have a lower throughput. This is expected since both nodes have to share the wireless channel. These results show that multiple devices can share conductive surfaces using carrier sense.}

\section{Future Work And Conclusion}
\label{sec:conc}
We introduce \name, a novel primitive that enables  small devices to communicate with each other using MIMO by using conductive surfaces. To do this, we identify two ways to make everyday surfaces compatible with wireless communication: coating them with conductive paint and using conductive cloth over tabletops. We also demonstrate the capability to use these conductive surfaces for Gbps data rates. We outline limitations and avenues for future work.

{\bf Security.} Our work focuses on characterizing conductive surfaces for the purposes of MIMO and high throughput communication. Our results however show that at the ISM bands, devices on the surface have a 15--25~dB advantage over those in air. We can in principle use this SNR advantage to design physical layer coding mechanisms that can enable physical layer security. Further, the multipath properties over the surface are distinctively different from in-air transmissions that could also be used for creating secret keys between the two devices. Further, we can transmit over a wideband of frequencies and perform frequency hopping. Leveraging this to design a secure communication system would be an interesting future direction.

{\bf  Multi-band surface ASIC.} Our Gbps platform is limited to using multiple off-the-shelf Wi-Fi chipsets. Designing a single chip across all these frequencies is technically feasible and would be a worthwhile engineering exercise.

{\bf Custom designed materials.}
Our system uses off-the-shelf conductive spraypaints and cloths that are designed for electromagnetic shielding and not for communication. However, there is nothing that fundamentally limits us from designing custom conductive surfaces with better properties for communication.  With custom designed spray paints and cloths we could decrease the amount of signal attenuation over distance and further limit the amount of signal that radiates into the environment. Further as these \red{cloths} and paints come in different colors they can be better integrated.

\section{Acknowledgements}
The researchers were funded by awards from the National Science Foundation, Sloan fellowship and the UW Reality Lab, Facebook, Google and Huawei.

\bibliographystyle{acm}
\bibliography{inkcomm} 

\balance{}
\end{document}